\documentclass[journal]{new-aiaa}

\usepackage{microtype}
\usepackage{amsmath}

\usepackage{amssymb}
\usepackage{gensymb}
\usepackage{graphicx}
\usepackage{pgf}
\usepackage{float}
\usepackage{ifdraft}
\usepackage[hyphens]{url}
\usepackage{hyperref}
\usepackage{enumitem}

\usepackage{eqexpl}
\eqexplSetIntro{where:} 
\eqexplSetDelim{=} 

\usepackage{ar}

\usepackage{multicol}

\usepackage{siunitx}
\usepackage{longtable}
\usepackage{tabularx}
\usepackage{booktabs}
\usepackage{multirow}
\usepackage{tablefootnote}
\usepackage{tabularray}
\UseTblrLibrary{booktabs}
\UseTblrLibrary{counter}

\usepackage{tikz}
\usetikzlibrary{positioning}
\usetikzlibrary{shapes.geometric}
\usetikzlibrary{shapes.arrows}
\usetikzlibrary{decorations.pathmorphing, decorations.pathreplacing, calc}
\usetikzlibrary{arrows.meta, positioning}

\usepackage{tikz-cd}
\tikzcdset{
    math mode=false
}

\usepackage{xcolor}

\definecolor{myorange}{RGB}{255, 165, 0}
\definecolor{mydarkseagreen}{RGB}{143, 188, 143}
\definecolor{mydodgerblue}{RGB}{30, 144, 255}

\colorlet{b}{red!13!white}
\colorlet{m}{yellow!20!white}
\colorlet{g}{green!18!white}

\usepackage{titlesec}

\usepackage{caption}
\usepackage[numbers,sort&compress]{natbib}

\usepackage{adjustbox}

\usepackage[titletoc,title]{appendix}

\usepackage[inkscapearea=page]{svg}

\usepackage{subcaption}

\usepackage{afterpage}
\usepackage[section]{placeins}

\usepackage{nth}

\usepackage{xstring}

\usepackage{authblk}

\usepackage{todonotes}
\setuptodonotes{inline,backgroundcolor=yellow!20,prepend,caption={TODO}}

\newcommand{\Rey}{\rm Re}

\newcommand{\Cp}{C_p}
\newcommand{\Cpo}{C_{p_0}}
\newcommand{\Cpm}{C_{p_{\rm min}}}
\newcommand{\Cpom}{C_{p_{0,\rm min}}}
\newcommand{\Cpcr}{C_{p_{\rm crit}}}
\newcommand{\Mcr}{M_{\rm crit}}
\newcommand{\Mdd}{M_{\rm dd}}
\newcommand{\Mi}{M_\infty}
\newcommand{\Ncr}{N_{\rm crit}}

\title{NeuralFoil: An Airfoil Aerodynamics Analysis Tool Using Physics-Informed Machine Learning}

\author{Peter Sharpe\footnote{PhD Candidate, AIAA Student Member} and R. John Hansman\footnote{T. Wilson Professor in Aeronautics, AIAA Fellow}}
\affil{Massachusetts Institute of Technology, Cambridge, MA}

\begin{document}

\maketitle

\begin{abstract}

    \emph{NeuralFoil} is an open-source Python-based tool for rapid aerodynamics analysis of airfoils, similar in purpose to XFoil. Speedups ranging from 8x to 1,000x over XFoil are demonstrated, after controlling for equivalent accuracy. NeuralFoil computes both global and local quantities (lift, drag, velocity distribution, etc.) over a broad input space, including: an 18-dimensional space of airfoil shapes, possibly including control deflections; a $360\degree$ range of angles of attack; Reynolds numbers from $10^2$ to $10^{10}$; subsonic flows up to the transonic drag rise; and with varying turbulence parameters. Results match those of XFoil closely: the mean relative error of drag is 0.37\% on simple cases, and remains as low as 2.0\% on a test dataset with numerous post-stall and transitional cases. NeuralFoil facilitates gradient-based design optimization, due to its $C^\infty$-continuous solutions, automatic-differentiation-compatibility, and bounded computational cost without non-convergence issues.

    NeuralFoil is a hybrid of physics-informed machine learning techniques and analytical models. Here, physics information includes symmetries that are structurally embedded into the model architecture, feature engineering using domain knowledge, and guaranteed extrapolation to known limit cases. This work also introduces a new approach for surrogate model uncertainty quantification that enables robust design optimization.

    This work discusses the methodology and performance of NeuralFoil with several case studies, including a practical airfoil design optimization study including both aerodynamic and non-aerodynamic constraints. Here, NeuralFoil optimization is able to produce airfoils nearly identical in performance and shape to expert-designed airfoils within seconds; these computationally-optimized airfoils provide a useful starting point for further expert refinement.

\end{abstract}

\section{Nomenclature}

 {\renewcommand\arraystretch{1.0}
  \noindent\begin{longtable*}{@{}l @{\quad=\quad} p{5in}@{}}
      $\beta$ & Prandtl-Glauert compressibility correction factor, defined as $\sqrt{1 - \Mi^2}$ \\
      BL & boundary layer \\
      CFD & computational fluid dynamics \\
      $C_D$ & drag coefficient \\
      $C_L$ & lift coefficient \\
      $C_M$ & moment coefficient \\
      $\Cp$ & local pressure coefficient (in the real compressible flow) \\
      $\Cpo$ & local pressure coefficient in the equivalent incompressible flow \\
      $\Cpm$ & minimum local pressure coefficient (in the real compressible flow) \\
      $\Cpom$ & minimum local pressure coefficient in the equivalent incompressible flow \\
      $c$ & airfoil chord \\
      $c_\mathcal{D}$ & dissipation coefficient \\
      $c_f$ & skin friction coefficient \\
      $c_{\rm lower}$ & Kulfan (CST) shape parameters associated with the airfoil's lower surface \\
      $c_{\rm upper}$ & Kulfan (CST) shape parameters associated with the airfoil's upper surface \\
      $\gamma$ & ratio of specific heats of the working fluid; 1.4 for air near standard conditions \\
      $H$ & BL shape parameter, defined as $\delta^*/\theta$ \\
      $H^*$ & BL kinetic energy shape parameter, defined as $\theta^*/\theta$ \\
      LE & airfoil leading edge \\
      $M$ & local Mach number \\
      $\Mi$ & freestream Mach number \\
      $\Mcr$ & critical Mach number, defined as the lowest freestream Mach number where any point in the flow is supersonic \\
      $\Mdd$ & drag-divergent Mach number, defined as the freestream Mach number above which drag begins to rise rapidly ($\partial(C_D)/\partial \Mi \geq 0.1$) \\
      $\nu$ & kinematic viscosity \\
      $\Ncr$ & BL critical amplification factor (affected by surface roughness, freestream turbulence) \\
      RANS & Reynolds-averaged Navier-Stokes equations \\
      $\Rey_c$ & chord Reynolds number, defined as $u_\infty c / \nu$ \\
      $\Rey_\theta$ & momentum-thickness Reynolds number, a local BL quantity defined as $u_e \theta / \nu$ \\
      $s$ & coordinate along the airfoil surface, starting at the TE and proceeding over the top surface \\
      $\theta$ & BL momentum thickness \\
      $\theta_{\rm TE}$ & Trailing-edge angle \\
      TE & airfoil trailing edge \\
      $u_e$ & BL edge velocity \\
      $u_\infty$ & freestream velocity magnitude \\
      $u_{\rm max}$ & maximum velocity magnitude at any point in the flow field \\
      $x_{\rm tr}$ & actual laminar-turbulent transition location, relative to the leading edge; appropriate suffixes denote top- and bottom-surface measures \\
      $x_{\rm tr, forced}$ & location of any forced trips for bypass transition (e.g., turbulators, rivets), if present; $x_{\rm tr, forced} / c = 1$ if absent\\
      $x / c$ & nondimensional distance along the airfoil chord ($\text{LE} \rightarrow 0$, $\text{TE} \rightarrow 1$) \\
      $y / c$ & nondimensional distance normal to the airfoil chord (LE and TE at 0, barring any control surface deflections) \\
      $z_{\rm in}$ & input latent space vector \\
      $z_{\rm out}$ & output latent space vector \\
  \end{longtable*}}

\section{Introduction}

In conceptual aircraft design, the problem of shaping a typical wing is usually decomposed into two parts: planform design and airfoil design. The latter, which is the focus of this work, is a multidisciplinary design problem that requires consideration of a variety of aerodynamic, structural, and manufacturing objectives and constraints. A non-exhaustive list of major considerations could include:
\begin{itemize}
    \item Profile drag across the expected operating range of the airfoil (spanning lift coefficients, Reynolds numbers, and Mach numbers), including adequate off-design performance \cite{drelaProsConsAirfoil1998};
    \item Pitching moment and aft-camber coefficients, which can drive tail sizing (modifying trim drag), affect divergence speed;
    \item Hinge moments and control effectiveness of any control surfaces, which drive actuator design and weight;
    \item Stall behavior, which can affect handling qualities and safety;
    \item Thickness at various points, in order to accommodate fuel volume and required structural members to resist failure (e.g., by bending, buckling, divergence, flutter, or control reversal);\cite{sharpeTaileronsAeroelasticStability2023}
    \item Sensitivity to boundary layer performance, freestream turbulence, and trips, all of which impose constraints on surface finish, cleanliness, and manufacturing tolerances \cite{eleshaky1993airfoil, seligHighLiftLowReynolds1997, liebeck1973class};
    \item Peak suction pressures, which affect the critical Mach number in transonic applications or cavitation in hydrodynamic applications;
    \item Shock stability and buffet considerations in transonic applications;
    \item Manufacturability, which might include flat-bottom airfoil sections, strictly-convex airfoil shapes (e.g., to accommodate shrink-coverings, which are common in ultra-lightweight applications \cite{drelaLowReynoldsnumberAirfoilDesign1988}), or restrictions on trailing-edge angle.
\end{itemize}

These airfoil design drivers often differ considerably at different locations along the span of the wing, which often leads to a family of airfoils being used in the design of a given wing. To fulfill such design requirements, a designer will typically either find an existing airfoil or design a new airfoil. Given the specificity of the requirements illustrated in the list above, designing bespoke airfoils can yield considerable performance improvement.

Currently, three approaches are commonly used to computationally design new airfoils: inverse design methods, direct manual methods, and optimization methods. In the inverse design approach, popularized by Drela's XFoil code \cite{drelaXFOILAnalysisDesign1989} and Eppler's Profil code \cite{profil, tao_bs_thesis}, conformal mapping methods are used to reconstruct a new airfoil shape from a user-specified pressure distribution. This has the benefit of allowing the engineer to directly operate on the most relevant aerodynamic quantities, and it produces considerable design insight. However, it can be time-consuming to produce airfoils that satisfy non-aerodynamic constraints (e.g., manufacturability, spar thickness) due to the lack of direct control here. Conformal mapping methods are also only strictly applicable to potential-flow-governed regions, so the specified pressure distribution is often only the inviscid, rather than the viscous (true) pressure distribution\footnote{This can be alleviated by using nonlinear optimization to target the viscous pressure distribution, albeit with reduced numerical robustness.}.

In the direct manual method (or ``geometric design'' method, using parlance from XFoil), an engineer formulates an airfoil shape directly and modifies this iteratively by hand. Aerodynamic analysis is performed by any code that will perform the forward problem (geometry $\rightarrow$ aerodynamics), such as XFoil, MSES \cite{mses}, or any RANS-based CFD code \cite{adlerCFDNotCFD2022}. This allows for easier satisfaction of non-aerodynamic constraints. However, it is predictably more difficult to directly target aerodynamic quantities. Significant user expertise is also required to identify the most relevant geometric parameters and to make effective changes to the airfoil shape.

In the optimization approach, a parameterized airfoil shape is optimized to minimize a cost function and satisfy specified constraints (which may be both aerodynamic and non-aerodynamic). At first glance, this appears to be an automation of the direct manual method. However, Drela and others note the surprising difficulty of posing the correct optimization problem \cite{drelaProsConsAirfoil1998,krooMultidisciplinaryOptimizationApplications1997}, so this approach often requires just as much (if not more) human expertise than the direct manual method. As stated by Drela \cite{drelaProsConsAirfoil1998}, optimization-based airfoil design ``is still an iterative cut-and-try undertaking. But compared to [direct] techniques, the cutting-and-trying is not on the geometry, but rather on the precise formulation of the optimization problem.'' To support this, Drela gives compelling case studies of how airfoil design optimization can go awry in the absence of user review and care.\footnote{Some codes, like LINDOP \cite{mses}, alleviate this somewhat by using a hybrid of the direct and optimization approaches: update directions are computed by an optimizer, but the actual changes are reviewed and implemented by a human between iterations.}However, despite these cautionary notes, the optimization-based airfoil design approach also offers compelling benefits: resulting airfoil performance can equal that of airfoils by an expert \cite{drelaProsConsAirfoil1998}, particularly on problems with unique or otherwise non-intuitive constraints. This optimization process can also require orders-of-magnitude less engineering time, and it provides a systematic and disciplined approach that is especially suited to the most challenging design problems \cite{he2019robust}.

In all these methods, some form of a computational tool for airfoil aerodynamics analysis is required. For subsonic airfoils, the gold standard of such tools is XFoil \cite{drelaXFOILAnalysisDesign1989}. Morgado et al. find that XFoil is more accurate than RANS-CFD-based tools in this regime \cite{morgado2016xfoil}, yet it has a computational cost that is roughly 1,000x lower than RANS approaches -- a testament to the power of its modeling approach, which strongly-couples integral boundary layer and potential flow methods. A complete description of this modeling approach is available in Drela's \textit{Aerodynamics of Viscous Fluids} \cite{drelaAerodynamicsViscousFluids2019}, and in recent state-of-the-art work by Zhang \cite{zhangThreedimensionalIntegralBoundary2022,zhangNonparametricDiscontinuousGalerkin2017}. However, despite XFoil's many strengths, it has several attributes that make it less-than-ideal for directly driving numerical optimization studies \cite{adlerCFDNotCFD2022}. Among these:

\begin{itemize}
    \item XFoil is not guaranteed to produce a solution. When an ``ambitious'' calculation is attempted, XFoil often fails to provide a converged solution; the unconverged result often has wildly-diverging values and is effectively unusable. In some cases, calculations can lead to infinite loops or process crashes due to unhandled exceptions. While this is acceptable in certain applications (e.g., manual direct analysis), it is generally unacceptable for use in numerical optimization. Instead, optimization strongly benefits from a robust analysis tool that always produces a result, even if that result has degraded accuracy; this allows the analysis to steer the optimizer back towards the design space of reasonable airfoils \cite{he2019robust}. A particularly useful attribute is when the model is deliberately made to be slightly pessimistic (e.g., overestimate drag) in regions of the design space with high uncertainty, adding further optimization pressure towards reasonable designs with low performance uncertainty.
          \begin{itemize}
              \item More generally, design optimization is not the only application that strongly benefits from an aerodynamics analysis tool that always produces an answer. Other examples where a non-answer, infinite loop, or crashed process are unacceptable include real-time control (e.g., as an aerodynamic model for a model-predictive controller onboard an aircraft) and flight simulation.
          \end{itemize}
    \item XFoil solutions are not necessarily unique, and slightly different solutions can be obtained for the same analysis problem (airfoil shape, angle of attack, and Reynolds and Mach numbers). In practice, this manifests as an effective hysteresis depending on whether the angle of attack is swept up or down. This flow non-uniqueness is in fact a real physical\footnote{For example, flow over an airfoil may separate as its angle of attack increases past $12\degree$, but it may not fully reattach until the angle of attack descends back to below $11\degree$} effect \cite{jamesonAirfoilsAdmittingNonunique1991,kuzmin2012non,he2019robust}. However, this non-uniqueness can be exceptionally problematic for numerical optimization, as an infinitesimal change in an input parameter can result in the solution jumping to a different Newton basis of attraction. Therefore, there is no limit to how sensitive performance can be to input parameters, which hampers techniques like finite-differencing for gradient-based optimization.
    \item XFoil solutions are non-smooth\footnote{precisely, they are $C^0$ continuous but not $C^1$ continuous} with respect to input parameters, which makes them fundamentally incompatible with gradient-based optimization. (Any attempt to directly optimize XFoil results with gradient-based methods invariably results in premature stopping at a local minimum.) Interestingly, this is a consequence of how laminar-turbulent transition is handled by XFoil's integral boundary layer solver. This solve requires the use of laminar and turbulent boundary layer \emph{closure models}, which are curve-fitted functions that yield various necessary quantities ($H^*$, $c_f$, $c_\mathcal{D}$, etc.) of the von Karman integral momentum and kinetic energy equations as a function of the two values that parameterize the boundary layer ($H$, $\Rey_\theta$). The laminar and turbulent versions of these functions differ. XFoil implements a cut-cell approach on the transitioning interval, which restores $C^0$-continuity (i.e., transition won't truly ``jump'' from one node to another discretely); however, a sharp change in gradient occurs whenever an individual node switches its equation from laminar to turbulent. Adler et al. provide a graphical depiction of this phenomenon \cite{adlerCFDNotCFD2022}.
    \item Most interfaces between an optimizer and XFoil communicate through a series of text files (i.e., hard disk), rather than by sharing data in memory (i.e., RAM). Given the quick speed of an individual XFoil run, this input-output overhead imposes a non-negligible performance penalty. While this programming-language-agnostic interface is arguably one of the reasons for XFoil's long-enduring popularity, it comes at the cost of runtime performance if the tool is to be used in a high-throughput setting (e.g., for design optimization or aerodynamic database construction).
\end{itemize}

This motivates the development of a new airfoil aerodynamics analysis tool that captures the advantages of XFoil (accuracy, speed) while mitigating these drawbacks (i.e., incompatibility with gradient-based optimization, non-convergence challenges). In recent years, many fields have benefited from a hybrid data-and-theory approach \cite{bruntonDataDrivenScience2017}, where data-driven models are used to augment traditional physics-based models with learned closures. This work presents a similar physics-informed approach as applied to analyzing airfoil aerodynamics.

\section{NeuralFoil Tool Description and Methodology}
\label{sec:methodology}

\subsection{Overview}

\begin{figure}[H]
    \centering
    \begin{tikzpicture}[
            node distance=0.7 cm and 0.7 cm,
            auto,
            box/.style={
                    rectangle,
                    rounded corners,
                    draw=#1!50!black,
                    fill=#1!20,
                    thick,
                    text width=6cm,
                    align=center,
                    minimum height=1cm
                },
            title/.style={font=\bfseries},
            list/.style={align=left}
        ]

        \node[box=myorange] (inputs) {
            \textbf{Inputs}
            \begin{itemize}
                \item Airfoil shape, parameterized as described in Section \ref{sec:airfoil-parameterization}
                \item Angle of attack $\alpha$
                \item Reynolds number $\Rey_c$
                \item Mach number $\Mi$
                \item Freestream turbulence $\Ncr$
                \item Forced trips $x_{\rm tr, top} / c$, $x_{\rm tr, bot} / c$ (optional)
                \item Control surfaces (both hinge points and deflection angles)
            \end{itemize}
        };

        \node[box=mydarkseagreen, right=of inputs, text width=2.5cm] (neuralfoil) {
            \textbf{NeuralFoil}
        };

        \node[box=mydodgerblue, right=of neuralfoil] (outputs) {
            \textbf{Outputs}
            \begin{itemize}
                \item Bulk outputs (scalars):
                      \begin{itemize}
                          \item Lift coefficient $C_L$
                          \item Drag coefficient $C_D$
                          \item Moment coefficient $C_M$
                          \item Critical Mach number $\Mcr$
                          \item Upper- and lower-surface turbulent transition locations
                      \end{itemize}
                \item Detailed outputs (vectors):
                      \begin{itemize}
                          \item BL Momentum thickness $\theta$
                          \item BL shape factor $H$
                          \item BL edge-velocity distribution\footnote{This contains identical information as the surface pressure distribution, under thin-shear-layer assumptions \cite{drelaXFOILAnalysisDesign1989}} $u_e/u_\infty$
                      \end{itemize}
            \end{itemize}
        };

        \draw[-Latex] (inputs) -- (neuralfoil);
        \draw[-Latex] (neuralfoil) -- (outputs);
    \end{tikzpicture}
    \caption{User-facing inputs and outputs of the NeuralFoil model.}
    \label{fig:neuralfoil_io}
\end{figure}
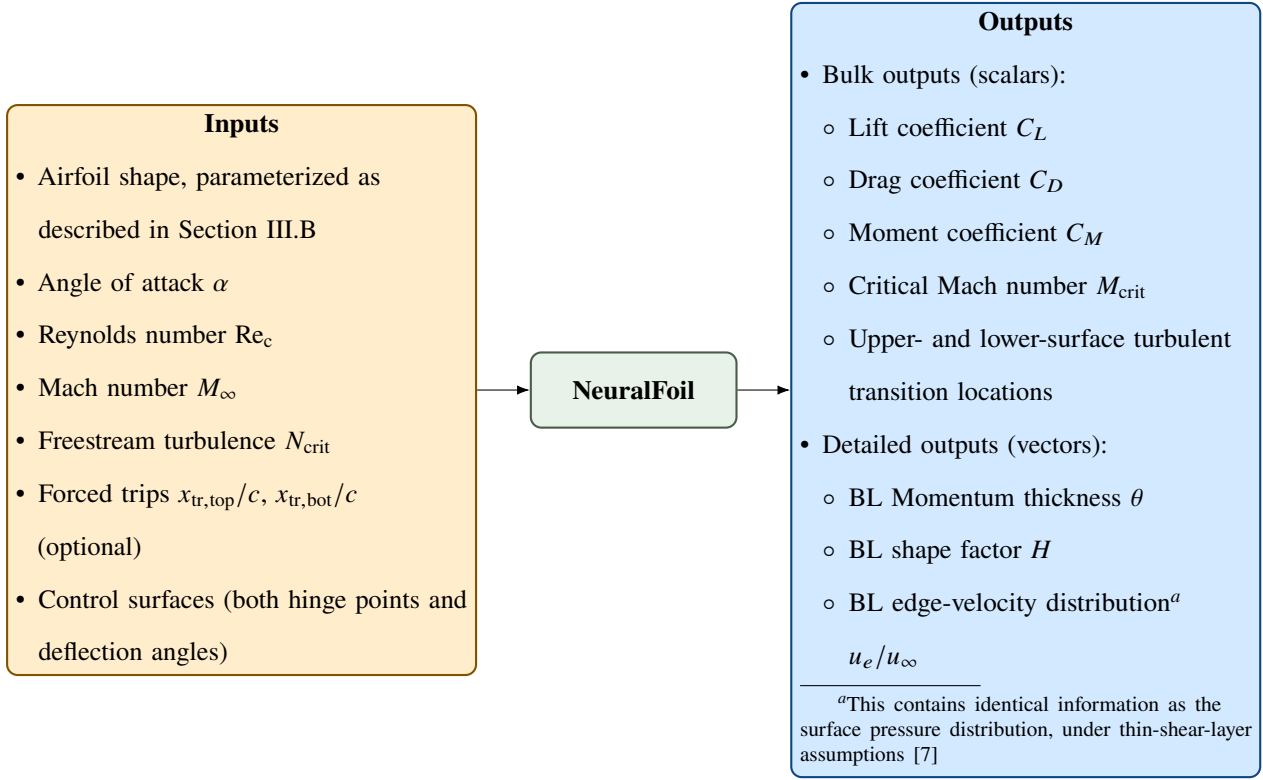

NeuralFoil is a tool for rapid aerodynamics analysis of airfoils, similar in purpose to XFoil \cite{drelaXFOILAnalysisDesign1989}. A precise list of inputs and outputs to NeuralFoil is given in Figure \ref{fig:neuralfoil_io}.

\noindent Although NeuralFoil's results will be most accurate in ``well-behaved'' flow regimes (e.g., attached flow), reasonable aerodynamics estimates can be expected:
\begin{itemize}
    \item For nearly all practical single-element airfoil shapes that can be analyzed with XFoil, including modifications for trailing-edge control surface deflections up to substantial deflections (roughly $\pm 40\degree$)
    \item Across the $360\degree$ angle of attack range, by leveraging analytical post-stall models regressed from high-$\alpha$ wind tunnel data by Truong \cite{truongAnalyticalModelAirfoil2020}
    \item Across a large range of Reynolds numbers (roughly $10^2$ to $10^{10}$; described in Table \ref{tab:flow_conditions}), with physically-sensible extrapolation even beyond this range (e.g., Stokes flow limit)
    \item At Mach numbers from zero to the transonic drag rise
\end{itemize}

\noindent NeuralFoil has a mathematical form that is fully explicit (i.e., no iterative solvers are used; there is no value-dependent code execution). This guarantees that a deterministic result is returned in bounded computational time and without any need for initial guesses. This also makes compatability with automatic differentiation frameworks much more straightforward, as the computational graph is static for any input. At a high level, the mathematical model within NeuralFoil consists of the following main steps:

\begin{enumerate}
    \item \textbf{Pre-solve}: Converts the user-supplied airfoil shape into the required parameterization (including any control surfaces, which are made part of the airfoil geometry as described in Section \ref{sec:control-surfaces})
    \item \textbf{Encoding}: Transforms all inputs (except Mach number, which is treated later) into an appropriate input vector space for the neural network. This is performed using prescribed functions based on domain knowledge.
    \item \textbf{Learned Model}: A learned neural network maps from this latent vector space to another latent vector space.
    \item \textbf{Decoding}: Transforms the outputs of this neural network back into the space of user-facing outputs. This is also performed using prescribed functions based on domain knowledge.
    \item \textbf{Uncertainty quantification, model fusion, and extrapolation}: Based on the neural network's self-reported trustworthiness (via a process described in Section \ref{sec:uq}), the network's results are merged with analytical models for airfoil behavior in massively-separated flow conditions.
    \item \textbf{Compressibility correction}: The solution is corrected for non-zero Mach numbers using analytical methods, as described in Section \ref{sec:compressibility}.
\end{enumerate}

\noindent In the following sections, we will describe this model architecture and theoretical basis in more detail. Readers primarily interested in general performance metrics and validation studies are invited to skip ahead to Section \ref{sec:results}. Readers primarily interested in a practical aerodynamic shape optimization example are directed to Section \ref{sec:optimization}.

\subsection{Pre-Solve}
\label{sec:airfoil-parameterization}

\subsubsection{Airfoil Geometry Parameterization}

NeuralFoil accepts user-specified airfoil shapes in a variety of common formats -- for example, as an array of $(x, y)$ coordinates, as a standard coordinate-array \texttt{*.dat} file, as a series of CST parameters, or as an Airfoil-class object within the AeroSandbox aircraft design optimization framework \cite{sharpeAeroSandboxDifferentiableFramework2021}.

\begin{figure}[h]
    \centering
    \adjustbox{trim=0cm 1.5cm 0cm 0cm}{
        \includesvg[width=0.65\textwidth]{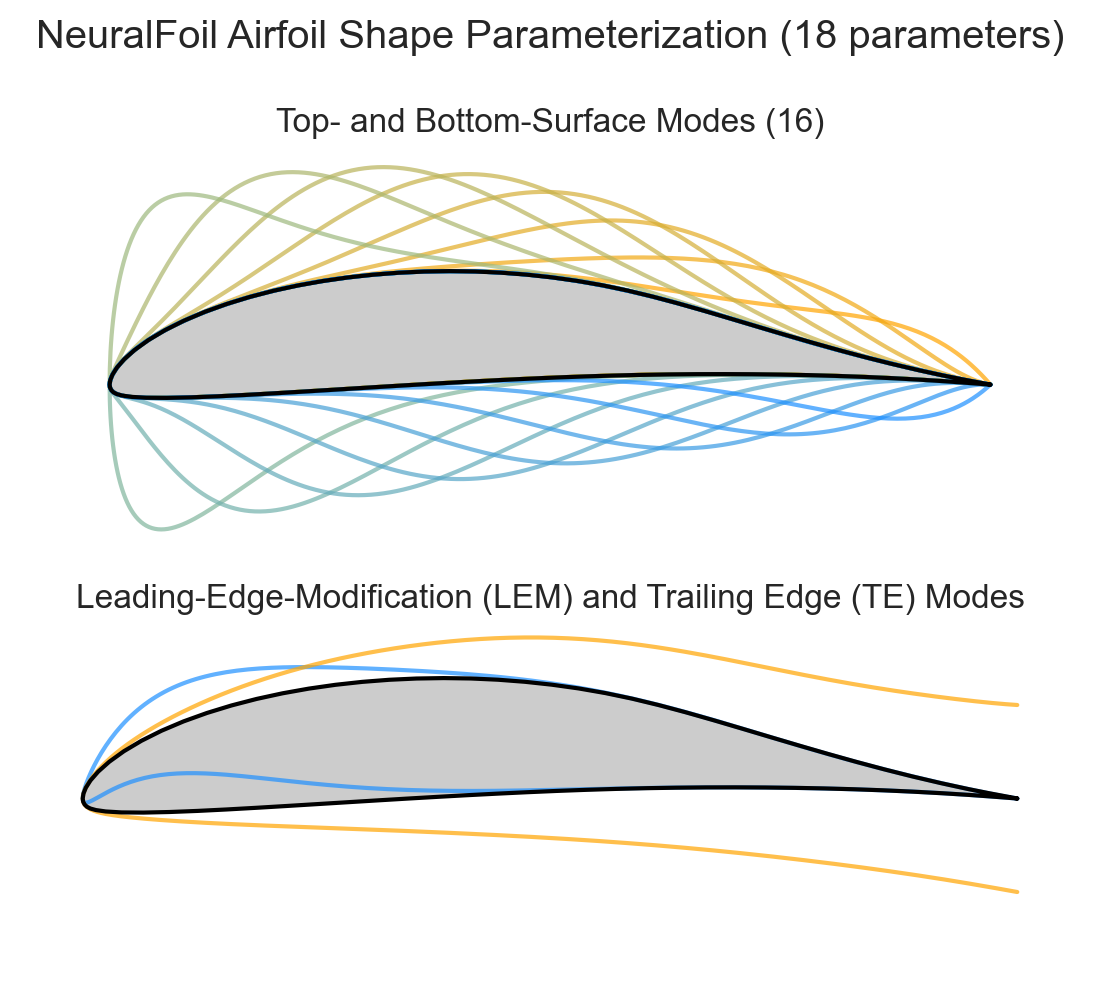}
    }
    \caption{Geometry input parameterization used by NeuralFoil. Parameterization is an 18-parameter CST (Kulfan) parameterization \cite{kulfanUniversalParametricGeometry2008,kulfanModificationCSTAirfoil2020, mastersGeometricComparisonAerofoil2017}. Each colored line in the figure represents a mode shape associated with one of these parameters; modes are linearly combined to form the airfoil shape.}
    \label{fig:neuralfoil_parameterization}
\end{figure}

Underneath this interface layer, NeuralFoil converts this specified airfoil geometry to an 8-parameter-per-side CST (Kulfan) parameterization, including Kulfan's added leading-edge-modification (LEM) and trailing-edge thickness parameter \cite{kulfanUniversalParametricGeometry2008,kulfanModificationCSTAirfoil2020}. This gives a total of 18 parameters to describe a given airfoil shape, which are illustrated in Figure \ref{fig:neuralfoil_parameterization}. This parameterization family was chosen due to work by Masters \cite{mastersGeometricComparisonAerofoil2017} and others, which shows that this is one of the most parameter-efficient representations of airfoil shape. Kulfan's parameterization is strongly related to an orthogonal polynomial decomposition using Bernstein polynomials. Because of this, the format is interpretable: it is a linear combination of mode shapes, each of which is roughly locally-supported\footnote{this contrasts with approaches such as taking an SVD over a standard airfoil database, which creates less-interpretable mode shapes}. Another benefit of the CST parameterization is interoperability, as it is commonly implemented in existing aerospace tools such as OpenVSP \cite{mcdonaldOpenVehicleSketch2022}.

The CST parameterization allows for varying numbers of degrees of freedom. An 18-parameter representation was chosen based on the work of Masters \cite{mastersGeometricComparisonAerofoil2017}, which shows that error in aerodynamic force prediction decreases substantially near this threshold. Thus, this parameterization strikes an acceptable balance between parameterization error and dimensionality. The 18-parameter representation also corresponds to one of the initially-proposed discretization levels proposed by Kulfan \cite{kulfanUniversalParametricGeometry2008} (labeled in this work as ``BPO8''), so this parameterization is a natural choice for compatibility with existing airfoil design tools.

Conversion of user-specified airfoils to this format for NeuralFoil to use is automatically and efficiently handled as a least-squares fitting problem.

\subsubsection{Control Surfaces}
\label{sec:control-surfaces}

\begin{figure}[h]
    \centering
    \includesvg[width=\textwidth]{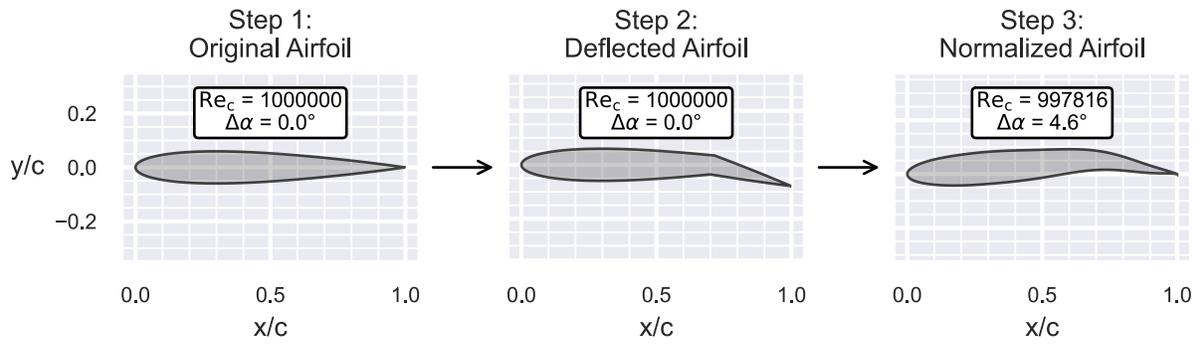}
    \caption{Illustration of the automatic procedure for handling control surface deflections in NeuralFoil.}
    \label{fig:control_surface_decomposition}
\end{figure}

Control surfaces are handled as a degenerate problem by re-normalizing the deflected airfoil shape. This is illustrated in Figure \ref{fig:control_surface_decomposition}. In step 1, an example airfoil is given. In step 2, a user-specified control surface deflection is applied. In step 3, the airfoil is re-normalized by applying the necessary similarity transformation (rotation, translation, and scaling) such that the leading-edge and trailing-edge locations are placed at their standard $(0, 0)$ and $(1, 0)$ locations in chord-normalized airfoil coordinates. The geometric rotation required for this re-normalization is later applied as a change in the effective angle of attack, $\Delta\alpha$. For consistency, the scaling factor required for this operation is also later used to scale the input Reynolds number appropriately, though the effect of this Reynolds scaling is typically minor. Finally, the resulting pitching moment must also be adjusted to account for the shifting of the force center due to translation during re-normalization.

Also visible in Figure \ref{fig:control_surface_decomposition} is the geometric effect of restricting the airfoil shape to the space of CST-parameterized airfoils, which is performed along with the re-normalization step between steps 2 and 3. The effect of this is that the sharp corners associated with the control surface deflection are smoothed out, which is a consequence of the smooth mode shapes associated with the CST parameterization. The rationale behind accepting this loss of fidelity here is that the control surface deflection invariably causes a turbulent transition at the deflection point on the suction side, regardless of whether the sharp or smoothed geometry is used. Because the turbulent boundary layer that follows the hinge is less sensitive to the pressure distribution (and hence, airfoil shape), the loss of geometric accuracy is less significant. Nevertheless, some loss of aerodynamic accuracy is to be expected.

\begin{figure}[H]
    \centering
    \includesvg[width=\textwidth]{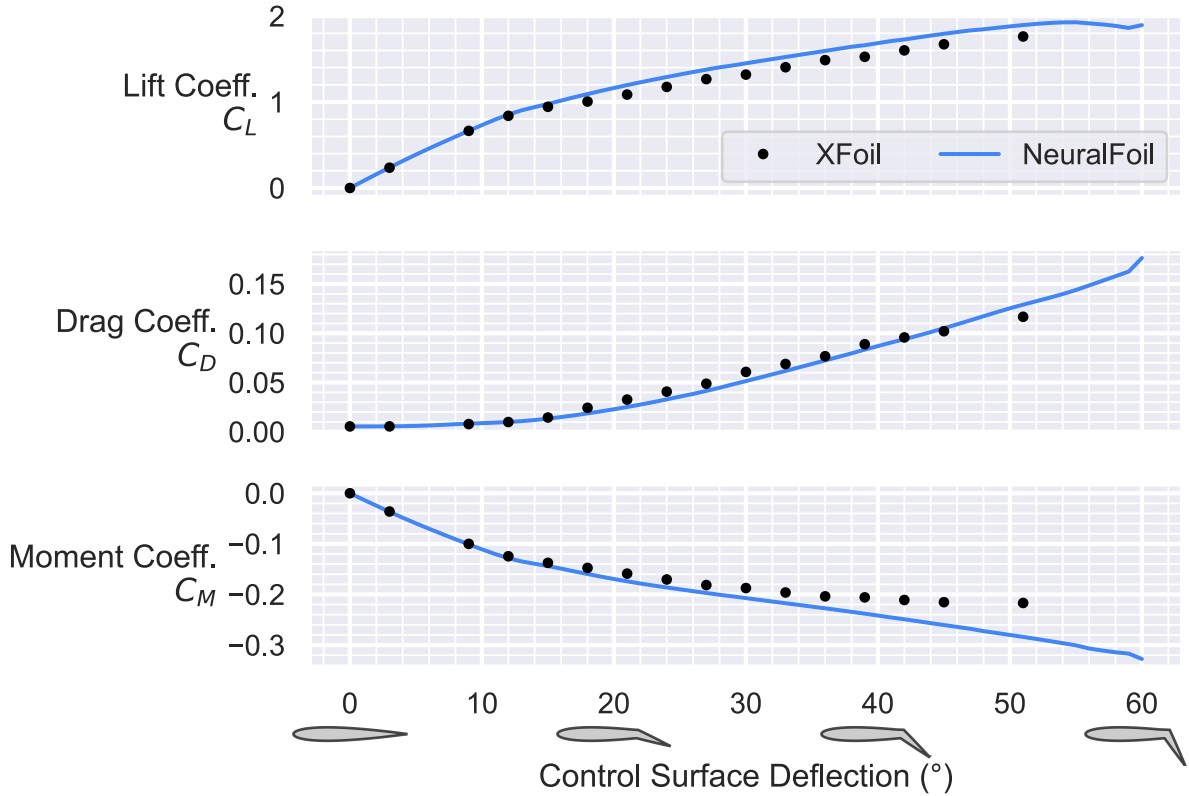}        \caption{Accuracy of NeuralFoil on an airfoil with large control surface deflections. Here, we show aerodynamics results from both NeuralFoil and XFoil. All runs are on a NACA0012 airfoil at $\Rey_c = 10^6$ and $\alpha = 0\degree$, with varying control surface deflections on a trailing-edge flap hinged at $x/c=0.70$.}
    \label{fig:control_surface_accuracy}
\end{figure}

To quantify this loss of accuracy, Figure \ref{fig:control_surface_accuracy} shows airfoil aerodynamics results obtained by both NeuralFoil and XFoil for various control surface deflections. Notably, reasonably close agreement is seen even for control surface deflections as aggressive as $\pm40 \degree$.

\subsection{Encoding, Learned Model Architecture, and Decoding}

After geometry parameterization, NeuralFoil transforms the user-facing inputs and outputs shown in Figure \ref{fig:neuralfoil_io} into intermediate vector spaces (latent spaces) that are more amenable to learning by a neural network. These latent spaces are carefully parameterized such that the learned model has a close-to-affine mapping of inputs to outputs. Effectively, this is feature engineering using domain-specific knowledge.

This encoding/decoding scheme has two major effects. First, it substantially increases the model's parameter efficiency\footnote{i.e., test-set accuracy, relative to the number of parameters (which represent model complexity and computational cost)}, since the training data has fewer nonlinearities in this latent space. Secondly, the combination of encoding functions, model architecture, and decoding functions can be used to guarantee physically-sound extrapolation beyond the dataset \cite{xuHowNeuralNetworks2021}. An example that illustrates both effects is the encoding of both the $\Rey_c$ input and the $C_D$ output into logspace. This means that an affine model in the latent space corresponds to a power-law $C_D(\Rey_c)$ model in user-facing space, which is a relationship supported by physical theory for both laminar and turbulent flows (e.g., Falkner-Skan and Schlichting boundary layer models \cite{drelaAerodynamicsViscousFluids2019}).

\subsubsection{Encoding}

The user-facing input space (shown in Figure \ref{fig:neuralfoil_io}) is transformed into the following input latent space, which is effectively what is seen by the neural network:

\begin{equation}
    z_{\rm in} = \text{Affine}\left( \begin{bmatrix}
            \text{Airfoil shape (18 parameters)} \\
            \sin(2 \alpha)                       \\
            \sin^2(\alpha)                       \\
            \cos(\alpha)                         \\
            \ln(\Rey_c)                          \\
            \Ncr                                 \\
            x_{\rm tr, top, forced}              \\
            x_{\rm tr, bot, forced}              \\
        \end{bmatrix} \right)
    \label{eq:encoding}
\end{equation}

The resulting $z_{\rm in}$ is 25-dimensional. Note that the inputs described in Equation \ref{eq:encoding} also undergo a simple elementwise affine transformation. This transformation is such that the distribution of typical inputs in the latent space has a mean of roughly zero and a standard deviation of roughly one\footnote{This normalization is performed on the basis of the training data, which is described later in Section \ref{sec:training-data}.}. Exact scaling and shift factors are available in the open-source codebase described in Section \ref{sec:reproducibility}. The purpose of this transformation is to improve the stability of the neural network training process, as well as to improve the performance of the weight-decay-based regularization strategy that is later used during training to improve generalization (described in Section \ref{sec:training-process}).

In addition to the previously-discussed log-space transformation of the Reynolds number, some nonlinear transformations on the angle of attack $\alpha$ are present and merit discussion as well. Encoding of the angle of attack into a purely-trigonometric representation embeds the periodic nature of the problem into the model architecture. For example, model evaluation at $\alpha=0\degree$ and $\alpha=360\degree$ are structurally identical, as these are encoded to the same location in the input latent space. The $\sin(2\alpha)$ and $\sin^2\alpha$ terms are directly proportional to common post-stall analytical models of lift and drag, respectively (such as those by Hoerner and Truong \cite{hoernerFluidDynamicLift1985, truongAnalyticalModelAirfoil2020}). The goal of this representation is to improve generalization performance in massively-separated flow conditions, where training data is scarce.

The freestream Mach number is a notable omission from the input latent space, and the learned model is both trained and evaluated on the equivalent incompressible flow. A compressibility correction is then performed after the neural network evaluation, as described in Section \ref{sec:compressibility}. This dimensionality reduction was performed to keep the required amount of training data more manageable, and because analytical models can accurately and quickly perform this compressibility correction (up to the critical Mach number).

\subsubsection{Learned Model Architecture}
\label{sec:learned-model-architecture}

After encoding inputs into an input latent space, NeuralFoil processes these inputs through a feedforward neural network (i.e., a multilayer perceptron). NeuralFoil offers eight different deep neural network models, which offer a tradeoff between accuracy and computational cost. This tradeoff is implemented as differences in the number and size of hidden layers, which are detailed in Table \ref{tab:model-sizes} for each model.

\begin{table}[H]
    \centering
    \caption{Neural network model sizes offered in NeuralFoil, offering a trade between accuracy and speed.}
    \label{tab:model-sizes}
    \begin{tabular}{lll}
        \toprule
        Model        & Hidden Layers & Neurons per Hidden Layer (width) \\ \midrule
        ``xxsmall''  & 1             & 48                               \\
        ``xsmall''   & 2             & 48                               \\
        ``small''    & 2             & 64                               \\
        ``medium''   & 3             & 64                               \\
        ``large''    & 3             & 128                              \\
        ``xlarge''   & 4             & 128                              \\
        ``xxlarge''  & 4             & 256                              \\
        ``xxxlarge'' & 5             & 512                              \\
        \bottomrule
    \end{tabular}
\end{table}

Each layer is a fully-connected linear layer. The activation function applied between each layer is the Swish function\footnote{sometimes written with an optional parameter $\beta$, which we set as $\beta=1$}, defined as $\sigma(x)=x / (1 + e^{-x})$. Compared to typical machine learning scenarios, the choice of activation function here has surprising importance, due to the desired properties of the resulting model. The Swish activation function is smooth ($C^\infty$-continuous), which makes the resulting neural network a smooth function as well. By preserving this continuity, NeuralFoil is made much more amenable to later gradient-based design optimization.

Likewise, the asymptotic behavior of the Swish function has key implications. Networks with activation functions that asymptote to piecewise-linear functions with differing positive and negative slopes (e.g., Swish, ReLU, LeakyReLU, Softplus) have fundamentally different extrapolation properties than networks with activation functions that asymptote to a constant (e.g., Sigmoid, Tanh). The former creates networks that asymptote to locally-affine functions, while the latter creates networks that asymptote to locally-constant functions. This observation about network extrapolation properties, discussed in greater detail by Xu et al. \cite{xuHowNeuralNetworks2021}, forms part of the justification for the latent space encodings described in the previous section where affine extrapolation is physically-consistent. Other activation functions were also considered. For example, the Softplus activation function meets many of the aforementioned criteria, but here it was abandoned in favor of Swish because networks with the latter achieved slightly better generalization performance.

Considerations during training of this network are discussed in Section \ref{sec:training-process}.

\subsubsection{Embedding of Angle of Attack Symmetry}

\begin{figure}[H]
    \centering
    \includesvg{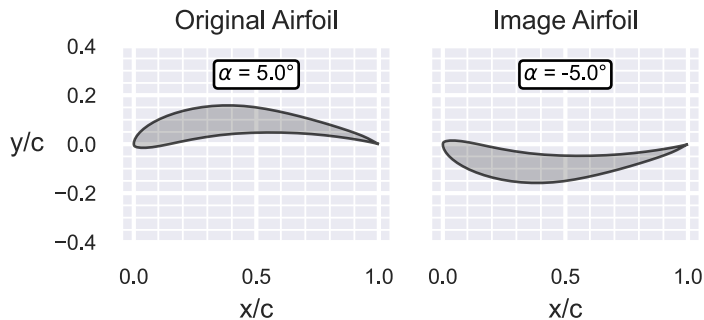}
    \caption{Illustration of the ``image'' approach used to structurally embed symmetry with respect to angle of attack in NeuralFoil.}
    \label{fig:alpha_symmetry}
\end{figure}

A key step during neural network evaluation (applied both during training and inference) is the embedding of physics symmetry with respect to angle of attack. The rationale behind this symmetry can be intuitively explained with the aid of the illustrations in Figure \ref{fig:alpha_symmetry}. Here, we consider a generic cambered airfoil at some angle of attack, as well as its ``image scenario'' consisting of the flipped airfoil at a flipped angle of attack. The flow physics in these two scenarios should be exactly equal and opposite, which will not generally be the case unless this is enforced. To enforce this symmetry, NeuralFoil computes the neural network output for both the original and flipped airfoil, and then merges the results. Outputs that are physically-guaranteed to exhibit even symmetry (e.g., $C_D$) are averaged, while outputs that exhibit odd symmetry (e.g., $C_L$, $C_M$) are subtracted. More details about this approach to embedding symmetries and invariants into neural networks are discussed in recent work by Zhang \cite{zhangThreedimensionalIntegralBoundary2022}.

This symmetry embedding has important impacts for end-use cases of NeuralFoil. For example, when NeuralFoil analyzes a symmetric airfoil at $\alpha=0\degree$, it will always yield a lift coefficient and pitching moment of \textit{exactly} zero\footnote{to within machine precision, with differences only due to compensated summation algorithms during matrix multiplication}, and upper- and lower-surface transition locations will also be exactly equal. This exactness would not be the case if other common methods, such as data duplication to learn physics symmetries, were used. A practical case where error in this symmetry would be particularly noticeable is on an aircraft's vertical stabilizer. Here, flows around symmetric airfoils at zero local incidence are common, and any error in enforcing this symmetry would lead to a noticeable nonzero net yaw moment on the aircraft.

\subsubsection{Decoding}

The raw result of the learned model is another latent-space vector, which is then transformed back into the user-facing output space. This decoding process is broadly motivated by the same considerations as the encoding process, where the goal is to embed physical intuition into the architecture. The output latent space consists of the following vector, shown here as a function of the user-facing outputs:

\begin{equation}
    z_{\rm out} = \text{Affine}\left( \begin{bmatrix}
            \text{Analysis Confidence}               \\
            C_L                                      \\
            \ln(C_D)                                 \\
            C_M                                      \\
            x_{\rm tr, top}                          \\
            x_{\rm tr, bot}                          \\
            \ln(\Rey_\theta) \text{ at 64 locations} \\
            \ln(H) \text{ at 64 locations}           \\
            u_e/u_\infty \text{ at 64 locations}     \\
        \end{bmatrix} \right)
    \label{eq:decoding}
\end{equation}

The output latent space has 195 dimensions, and, similar to the input latent space, undergoes an elementwise affine transformation to normalize the distribution of typical outputs. The ``analysis confidence'' output described in Equation \ref{eq:decoding} deserves special attention and is discussed in Section \ref{sec:uq}.

\begin{figure}[H]
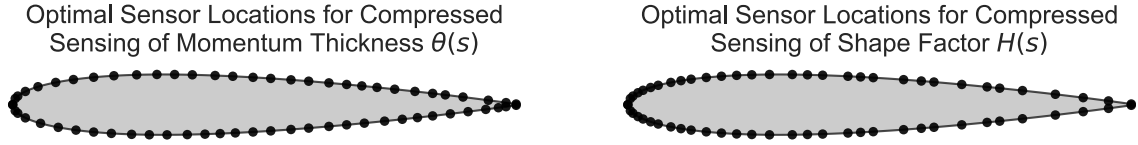

    \centering
    \begin{subfigure}[t]{0.49\textwidth}
        \centering
        \includesvg[width=\textwidth]{training/optimal_sensor_placement/optimal_sensors_theta.svg}
        \caption{Optimal sensor placement for minimum-error $\theta(s)$ reconstruction.}
    \end{subfigure}
    \begin{subfigure}[t]{0.49\textwidth}
        \centering
        \includesvg[width=\textwidth]{training/optimal_sensor_placement/optimal_sensors_H.svg}
        \caption{Optimal sensor placement for minimum-error $H(s)$ reconstruction.}
    \end{subfigure}
    \caption{Results from a compressed sensing study to determine optimal discrete locations to track boundary layer data for minimal reconstruction error.}
    \label{fig:optimal_sensor_placement}
\end{figure}

The boundary layer outputs in Equation \ref{eq:decoding} are computed at 64 locations along the airfoil surface, which are evenly spaced in the normalized $x$-coordinate on the top and bottom surfaces. These data points are conceptually similar to physical sensors, as they are effectively a discrete projection of the actual boundary layer flow, which itself is a continuous function of the surface coordinate $s$. The spacing of these discrete ``sensors'' was chosen on the basis of a compressed sensing study, the results of which are shown in Figure \ref{fig:optimal_sensor_placement}. In short, this study aimed to determine ``optimal sensor placement'' for airfoil boundary layer data: where should one place 64 discrete sensors on an airfoil to minimize the error when reconstructing the boundary layer data? Data for this study was generated from XFoil via the same process detailed in Section \ref{sec:training-data}. Compressed sensor placement was performed using approximate QR-factorization-based methods described by de Silva et al. \cite{de_Silva2021}

The results in Figure \ref{fig:optimal_sensor_placement} show that accurate reconstruction of the momentum thickness $\theta(s)$ is best achieved by roughly-uniform sensor placement. Interestingly, the optimal placement for the shape factor $H(s)$ is not uniform, but rather is concentrated near the leading edge. Since $H$ closely relates to laminar/turbulent behavior and transition location, this result is physically intuitive: small changes in $H$ near the leading edge have a disproportionate impact on the downstream flow.

\subsection{Uncertainty Quantification, Model Fusion, and Extrapolation}
\label{sec:uq}

The ``analysis confidence'' output detailed in Equation \ref{eq:decoding} can be loosely interpreted as a representation of NeuralFoil's self-reported model uncertainty. Quantitatively, this output is trained on binary features corresponding to whether an XFoil analysis with these input conditions converged. In effect, this is a classification problem based on whether the surrogate's underlying model returns trustworthy answers near a given point in the input space. The raw network output is a logit, which is then transformed into a confidence score in the range $(0, 1)$ using a logistic function. Example results of this ``analysis confidence'' metric are shown and discussed in Section \ref{sec:uq_results}.

Presenting the user with a measure of analysis confidence has several real-world benefits. First, it gives the user direct feedback on when the result of a requested analysis should be trusted. Surrogate models, especially those created with machine learning, invariably lose accuracy when extrapolating beyond the training data. This can be mitigated somewhat with embedded physics constraints, but ultimately NeuralFoil is no exception here. Furthermore, most neural surrogates give no obvious information about when this spurious extrapolation is occurring. By providing a confidence score, NeuralFoil can flag these scenarios; this may encourage a user to cross-check with a higher-fidelity model or pursue a less-uncertain design. Second, the confidence score can be used as a direct constraint during aerodynamic shape optimization driven by NeuralFoil. This enables a form of robust optimization where results are guaranteed to be within the region of the input space where the surrogate model is trustworthy.

Learning this ``analysis confidence'' binary classifier with no priors or physics knowledge presents a problem, however. Without such modifications, there is no guarantee that the model will extrapolate towards ``untrustworthiness'' (i.e., low analysis confidence) when far from the training data distribution. To enforce this behavior, the analysis confidence logit (i.e., logarithm of odds-ratio) is modified during both training and inference. The added term is the negative squared Mahalanobis distance\footnote{essentially, the Euclidian norm of the multidimensional generalization of the z-score} of the query point with respect to the training data distribution:

\begin{equation}
    \text{Analysis Confidence} = \sigma
    \left(
    \left(\text{Raw Logit}\right) -
    \left( \vec{z}_{\rm in} - \vec\mu \right) ^ T
    S ^{-1}
    \left( \vec{z}_{\rm in} - \vec\mu \right)
    \right)
    \label{eq:uq}
\end{equation}

\begin{eqexpl}[30mm]
    \item{Analysis Confidence} The final output of the analysis confidence model, in the range $(0, 1)$
    \item{$\sigma$} The logistic function, defined as $\sigma(x) = 1 / (1 + e^{-x})$
    \item{$\vec\mu$} The mean of the training data distribution in the input latent space
    \item{Raw Logit} The raw output of the neural network, before adding the Mahalanobis distance term
    \item{$S$} The covariance matrix of the training data distribution in the input latent space
    \item{$\vec{z}_{\rm in}$} The query point in the input latent space
\end{eqexpl}

Because the covariance matrix of the training data is positive-definite, the correction term is guaranteed to asymptote to $-\infty$ as the query point moves away from the training data in any direction. By contrast, the raw logit is structurally guaranteed to extrapolate to a locally-affine function, as previous discussed in Section \ref{sec:learned-model-architecture}. Therefore, this modification guarantees that the analysis confidence tends to zero far from the training data distribution. Because this modification is included at both train and inference time, this modification has minimal effect on the model's performance \textit{within} the training data distribution, and only serves to embed desirable extrapolation properties when \textit{outside} the distribution.

Another benefit of this modification is that it tends to make level sets of the analysis confidence more convex, since the modification is a quadratic form with a positive-definite Hessian\footnote{directly proportional to the inverse of the covariance matrix}. By regularizing the analysis confidence to be ``more convex'', this modification ultimately leads to better convergence in downstream gradient-based optimization problems that use the analysis confidence as a constraint.

The weighting of this Mahalanobis distance term could be varied, depending on how willing the developer of a neural surrogate is to allow the surrogate to extrapolate beyond the training distribution. In the case of NeuralFoil, a unit weighting (as shown in Equation \ref{eq:uq}) was chosen and appears to strike a good balance between extrapolation correctness and the amount of added nonlinearity. However, more broadly, this tradeoff forms an interesting area of further research in neural surrogates with self-reported trust metrics.

\begin{figure}[H]
    \centering
    \includesvg{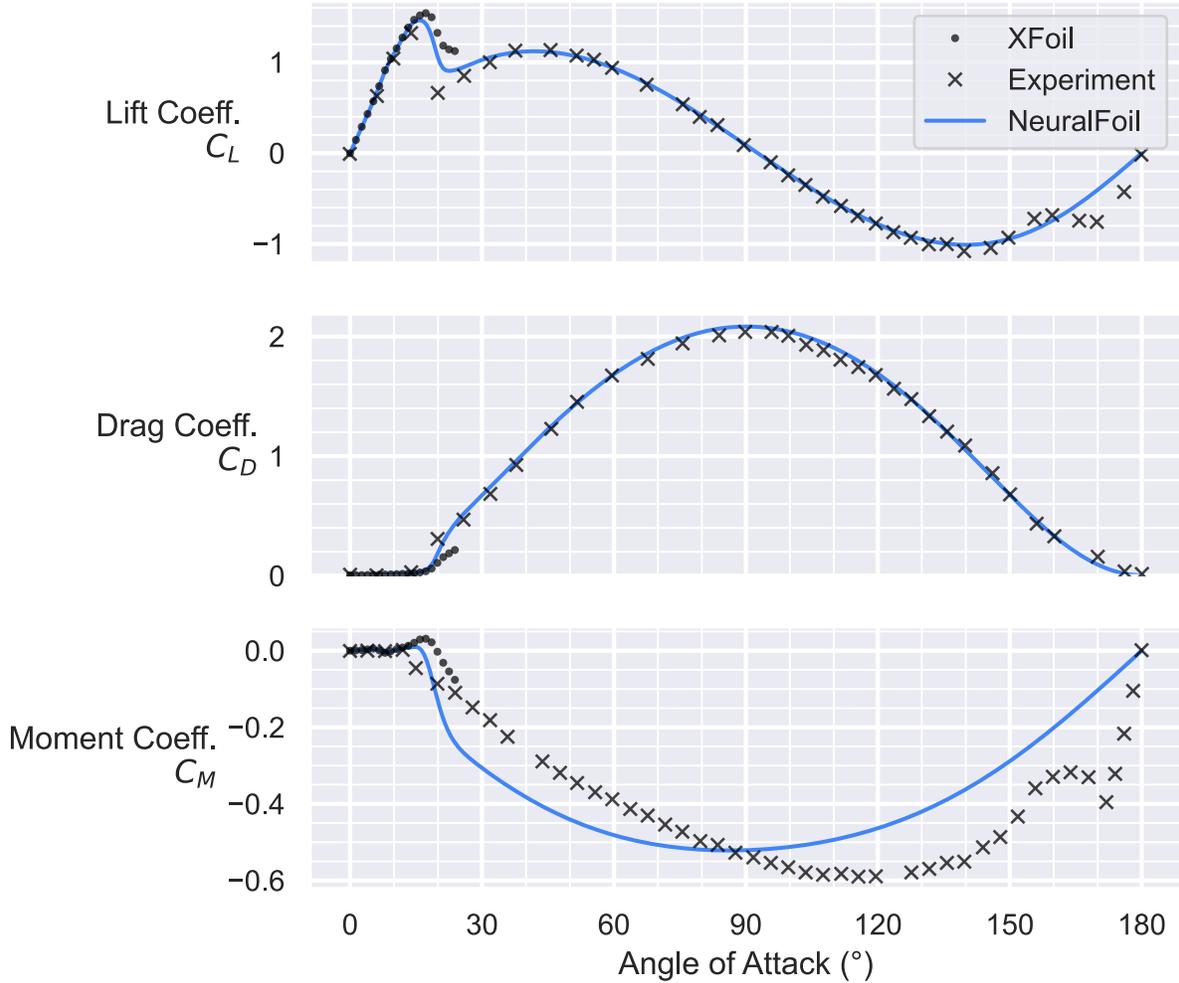}
    \caption{Illustration of the performance of NeuralFoil in massively-separated flow conditions. NACA0012 airfoil at $\Rey_c=1.8\times 10^6$. Experimental data reproduced from Langley wind tunnel data in NACA TN 3361 \cite{critzosAerodynamicCharacteristicsNACA1955}. NeuralFoil computes post-stall lift and drag quite accurately; moment computation is somewhat less accurate.}
    \label{fig:post_stall_extrapolation}
\end{figure}

Based on the analysis confidence of this learned model, NeuralFoil fuses its attached-flow results with empirical models for massively-separated (post-stall) flow conditions. Specifically, NeuralFoil uses the analytical post-stall models regressed by Truong \cite{truongAnalyticalModelAirfoil2020} to provide a more accurate prediction in these conditions. Example results showing this post-stall model fusion are given in Figure \ref{fig:post_stall_extrapolation}.

The NeuralFoil analysis results have some notable deviations from experiment. First, moment computation is somewhat less accurate than lift and drag computation. Secondly, NeuralFoil does not capture reversed-flow reattachment near $\alpha=180\degree$. Nevertheless, lift and drag results are relatively accurate, and results follow physically-sensible trends. In cases where post-stall data would be used (e.g., flight simulation, real-time control, helicopter blade stall, or recovering from a bad initial guess during design optimization), a less-accurate-but-still-reasonable answer may be useful. The model fusion shown in Figure \ref{fig:post_stall_extrapolation} is also smooth, facilitating later gradient-based optimization.

\subsubsection{Handling of Compressibility Effects}
\label{sec:compressibility}

At a high level, compressibility is handled by applying a correction factor to the incompressible results. This correction factor is computed using Laitone's rule \cite{laitoneNewCompressibilityCorrection1951}, which is a higher-order variant of the well-known Prandtl-Glauert and Karman-Tsien compressibility corrections. All of these compressibility corrections take two inputs: the pressure coefficient in the equivalent incompressible flow $\Cpo$ and the Mach number $M$; they return the pressure coefficient in the actual compressible flow $\Cp$. For comparison, all three corrections are given here, which conveniently show the higher-order terms retained in each successive relation:

\begin{align}
    \label{eq:prandtl-glauert} \text{Prandtl-Glauert:} \qquad        & \Cp = \frac{\Cpo}{\beta}                                                                                    \\
    \label{eq:karman-tsien} \text{Karman-Tsien:}\qquad               & \Cp = \frac{\Cpo}{\beta + \Mi^2 / (1 + \beta) \cdot \Cpo / 2}                                               \\
    \label{eq:laitone} \text{Laitone \cite{laitoneNewCompressibilityCorrection1951}:}\qquad & \Cp = \frac{\Cpo}{\beta + \Mi^2 / \beta \cdot \Cpo / 2 \cdot \left( 1 + \frac{\gamma - 1}{2} \Mi^2 \right)}
\end{align}

\noindent where $\beta = \sqrt{1- \Mi^2}$, and $\gamma$ is the ratio of specific heats (1.4 for air near standard conditions). In theory, these compressibility corrections should apply only to the pressure-derived forces on the airfoil, while the shear forces are relatively unaffected. To approximate this, NeuralFoil applies the compressibility correction to the lift force and pitching moment (which are pressure-dominated) but does not apply the correction to the drag force (which is often shear-dominated). This assumption, while relatively simple, proves to match compressible airfoil drag data computed with other methods quite closely, as shown in Section \ref{sec:validation_transonic}.

Another useful definition is that of the sonic pressure coefficient $\Cpcr$, which is the pressure coefficient below which flow goes supersonic. This is derived from the isentropic relations, yielding:

\begin{equation}
    \Cpcr = \frac{2}{\gamma \Mi^{2}} \left(
    \left(
    \frac{1 + \frac{\gamma - 1}{2} \Mi^{2}}{1 + \frac{\gamma - 1}{2}}
    \right)^{\frac{\gamma}{\gamma - 1}}
    - 1
    \right)
    \label{eq:sonic-pressure-coefficient}
\end{equation}

\begin{figure}[h]
    \centering
    \includesvg{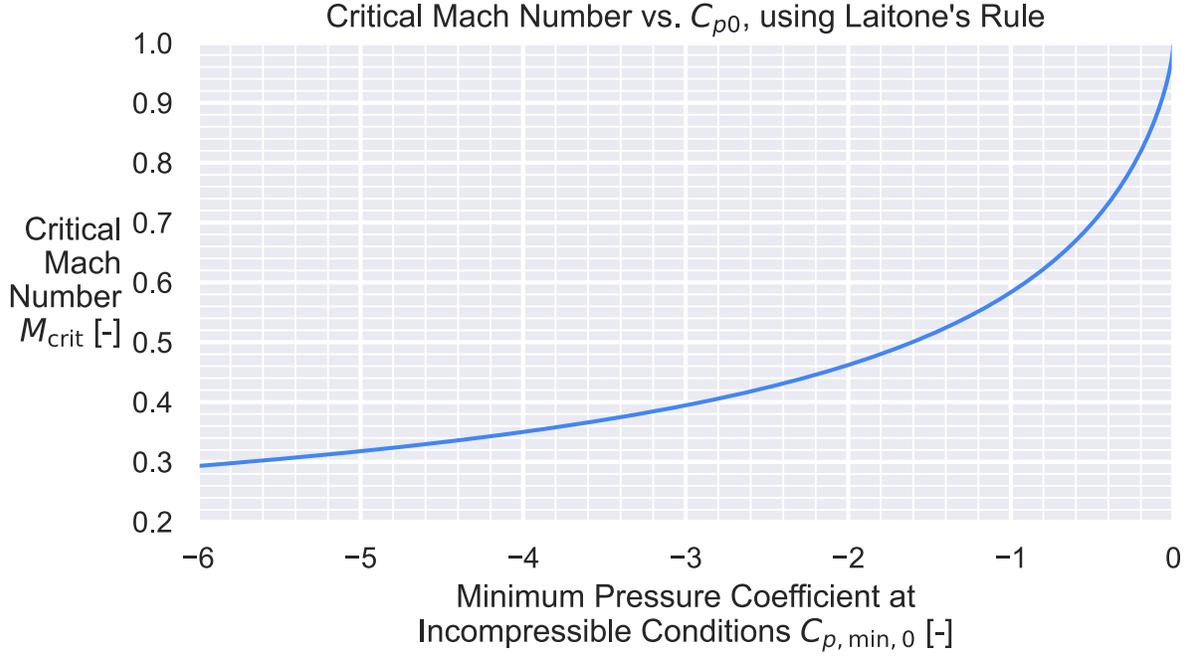}
    \caption{Laitone's rule allows a mapping from the minimum incompressible pressure coefficient $\Cpom$ to the critical Mach number $\Mcr$.}
    \label{fig:compressibility_corrections}
\end{figure}

At the location where supersonic flow first begins at $\Mcr$, Equations \ref{eq:laitone} and \ref{eq:sonic-pressure-coefficient} can be set equal. This yields a relation that maps the minimum value of the incompressible pressure coefficient $\Cpm$ to the critical Mach number $\Mcr$. This relation does not admit a closed-form solution, but it can be solved numerically; this is shown in Figure \ref{fig:compressibility_corrections}.

This implicit mapping via a nonlinear solve is less desirable at runtime, for two main reasons. First, this method is iterative and does not have a static computational graph, complicating compatibility with automatic differentiation tools. Secondly, it is not guaranteed to converge, depending on the initial guess. Instead, this relation can be replaced with an explicit surrogate model, which was obtained using symbolic regression (via PySR \cite{cranmerInterpretableMachineLearning2023}):

\begin{equation}
    \Mcr = \left(
    1.0083619
    - \Cpom
    + \left(-0.51058894 \cdot \Cpom \right)^{0.6553655}
    \right)^{-0.5536965}
    \label{eq:laitone_surrogate}
\end{equation}

Replacing the implicit mapping of Figure \ref{fig:compressibility_corrections} with the surrogate model of Equation \ref{eq:laitone_surrogate} introduces negligible error, with a corresponding mean RMS error in $\Mcr$ of $0.0014$ over a representative range of $\Cpom$ values\footnote{More precisely: finding the mean error with respect to Mach in the interval $M\in[0.001, 0.999]$.}. This relation allows NeuralFoil to estimate the critical Mach number $\Mcr$ relatively accurately using only incompressible quantities, as shown later in Section \ref{sec:validation_transonic} and Table \ref{tab:transonic_validation} relative to full-potential and RANS solutions. For the purposes of Equation \ref{eq:laitone_surrogate}, the incompressible $\Cpom$ is computed definitionally from surface values of velocity magnitude, as reported by the learned model:

\begin{equation}
    \Cpom = 1 - \left( \frac{u_{\rm max}}{u_\infty} \right)^2 \qquad \text{at }\ \Mi = 0
\end{equation}

Following a derivation from Mason \cite{masonTransonicAerodynamicsAirfoils2006}, an empirical relation for the shape of the drag rise beyond $\Mcr$ is included. Drag in this transonic regime, especially beyond the drag-divergent Mach number $\Mdd$, is relatively simple and typically errs on the side of over-estimating wave drag. Thus, NeuralFoil's predictions of \textit{when} transonic flow will occur are reasonably trustworthy, but wave drag results deep within this transonic regime are not. However, given that a primary goal of the NeuralFoil tool is to drive design optimization, this simple empirical model achieves its purpose of steering an optimizer away from thick transonic airfoils with strong shocks.

\subsection{Training Data Generation}
\label{sec:training-data}

Synthetic training data was generated by running XFoil on randomly-generated airfoil shapes and flow conditions. All training data is analyzed without compressibility in XFoil (i.e., $\Mi=0$); compressible effects were handled outside of the neural network training process, as described in Section \ref{sec:compressibility}. The stochastic procedure used to generate the airfoil shapes used in training can be described as follows:

\begin{enumerate}
    \item First, three parent airfoils are randomly selected from an airfoil database\footnote{This database consists of 2,174 airfoils, which are drawn from a variety of sources (most notably, the UIUC airfoil database \cite{uiuc_airfoil_database} and TraCFoil database \cite{etiembleTraCFoilFreePack2023}) and manually cleaned. The database is publicly available via AeroSandbox \cite{sharpeAeroSandboxDifferentiableFramework2021}.}, and converted to the Kulfan geometry parameterization. Three random weights are drawn, and the airfoils are merged into one based on these weights. This procedure effectively ensures that each training airfoil is derived from a unique combination of three parent airfoils.
    \item To increase the diversity of training data, several further modifications are applied:
          \begin{itemize}
              \item The thickness is randomly scaled by a factor drawn from $\operatorname{Lognormal}(\mu=0,\ \sigma=0.15)$.
              \item The airfoil's Kulfan parameters are perturbed by a random vector drawn from a multivariate normal distribution. The mean and covariance of this distribution are taken from the sample statistics of the airfoil database.
          \end{itemize}
    \item Flow conditions are randomly selected. Angle of attack $\alpha$ is drawn from the sum of a uniform and normal distribution, and Reynolds number $\Rey_c$ is drawn from a log-normal distribution. Distributional statistics for these variables and transition criteria are given in Table \ref{tab:flow_conditions}.
\end{enumerate}

\begin{table}[H]
    \caption{Summary statistics of flow conditions in the overall dataset, which is later partitioned into separate training and test datasets. Percentages refer to percentiles of the distribution. Strictly speaking, this table gives summary statistics for the subset of generated cases that were later associated with a successfully-converged XFoil run, so the true distribution of the generator is slightly wider. Suffixes `k', `M', and `T' denote $10^3$, $10^6$, and $10^{12}$, respectively.}
    \label{tab:flow_conditions}

    \centering
    \begin{tblr}{
            colspec={@{} llllllll@{}},
            row{1}={font=\bfseries},
        }
        \toprule
                                                    & Minimum                                                                                      & 2.5\%          & 25\%          & 50\% (Median) & 75\%          & 97.5\%         & Maximum        \\ \midrule
        Angle of attack $\alpha$                    & $-27.9\degree$                                                                               & $-17.3\degree$ & $-7.5\degree$ & $+0.9\degree$ & $+8.7\degree$ & $+17.6\degree$ & $+28.6\degree$ \\
        Reynolds number $\Rey_c$                    & $0.916$                                                                                      & $1.87$k        & $31.1$k       & $296$k        & $3.47$M       & $262$M         & $2.92$T        \\
        Critical amplification factor $\Ncr$        & \SetCell[c=7]{c}{Uniformly distributed in $[0, 18]$.}                                                                                                                                           \\
        Transition locations $x_{\rm tr, forced}/c$ & \SetCell[c=7]{c}{With 80\% probability, natural transition. Otherwise, uniform in $[0, 1]$.}                                                                                                    \\
        \bottomrule
    \end{tblr}
\end{table}

In total, 7,913,292 data points (i.e., airfoils and flow conditions) were generated with this procedure and analyzed with XFoil. Cases that did not result in converged XFoil solutions were not used to train physics outputs; however, this non-convergence was still recorded as a binary feature to train the analysis confidence output described in Section \ref{sec:uq}. Overall, $56\%$ of generated cases resulted in converged XFoil solutions, which suggests that this data-generating process does a reasonable job of spanning the space of inputs where XFoil convergence is likely. XFoil results that were purportedly converged but violated physical constraints (e.g., $\theta<0$, $H<1$, non-physical $u_e$ values) were also automatically filtered and treated as unconverged, though this was rare and affected only 0.03\% of cases. The distribution of training data points was sufficiently diverse that converged XFoil results were obtained with lift coefficients ranging from $-2.67$ to $+3.44$.

It is not currently known whether the number of points in this dataset is too little, too much, or appropriate. As a point of comparison, the number of data points used to train NeuralFoil is roughly two orders of magnitude larger than that of similar studies (see Table \ref{tab:ml_comparison}), although NeuralFoil's data distribution is substantially wider, and hence this may be warranted. In total, data generation required roughly 24 hours on MIT Supercloud, a computing cluster operated by the Lincoln Laboratory Supercomputing Center. It is possible either that similar performance could be achievable with much less data, or that more data could further improve performance. This is left as an area of future research.

\subsection{Training Process}
\label{sec:training-process}

The neural networks at the heart of this approach were trained with MIT Supercloud's GPU computing capabilities. The training process was implemented using the PyTorch \cite{paszkePyTorchImperativeStyle2019} framework. The RAdam optimizer \cite{liuVarianceAdaptiveLearning2019} was used with a decaying learning rate scheduled based on the plateau of the training loss.

Synthetic data generated using the procedure described in Section \ref{sec:training-data} was split into a training dataset (95\%) and a test dataset (5\%), with the latter used only to evaluate generalization accuracy. Critically, a small amount of weight decay (effectively, a $L^2$-norm penalty on all weight and bias parameters) was added to the loss function, which improves generalization performance and causes the network training to asymptote without over-fitting. Other techniques to improve generalization, such as batch normalization and dropout, were tested; however, it was found that weight decay alone consistently produced the models with the most accurate generalization to the test dataset. Various hyperparameters associated with training the model were optimized via a parallelized grid search. Because the weight decay regularization effectively limits the amount of overfitting that is possible, all models were trained until the test-set loss reached an asymptote.

The loss function used when training the network is a Huber loss metric on each of the latent-space outputs shown in Equation \ref{eq:decoding}. For this Huber loss, the $L^1$-to-$L^2$ transition point is at $\delta=0.05$ (recalling that one unit in the output latent space corresponds roughly to the training data's output standard deviation). The exception to this Huber loss metric is the analysis confidence output, which is instead converted to a binary cross-entropy loss. Exact training details are publicly-available as described in Section \ref{sec:reproducibility}.

Individual components of the loss function are weighted differently, to reflect differing importance during typical analysis and optimization. The highest weight is applied to drag, followed in descending order by lift, moment, and transition locations. The relative weightings of such parameters were optimized as a hyperparameter, with the goal of minimizing generalization error (i.e., test-set performance) of drag estimation. Interestingly, this error appears minimized not by tilting the weighting entirely towards drag, but by also adding small but significant weights to other outputs, such as the detailed boundary layer. Stated more informally, this suggests that the best way to compute drag is to learn a general model that estimates all physical outputs, rather than a model focused purely on drag. This suggests that the model learns deeper physics relationships than simple memorization, as the presence of intermediate representations results in improved generalization performance on bulk quantities.

\section{Results and Discussion}
\label{sec:results}

\subsection{Point Validation of NeuralFoil Accuracy with respect to XFoil}
\label{sec:validation_basic}

Qualitatively, NeuralFoil tracks XFoil very closely across a wide range of angles of attack and Reynolds numbers. In Figure \ref{fig:clcd_polar}, we compare the performance of NeuralFoil to XFoil on aerodynamic polar prediction. This study aims to evaluate the generalization performance: to what extent is the model learning physics vs. merely memorizing its training data? This generalization performance forms an important metric of usefulness for real-world design problems.

\begin{figure}[h]
    \centering
    \includesvg[width=\textwidth]{benchmarking/neuralfoil_point_comparison.svg}
    \caption{Generalization performance of NeuralFoil, assessed by sample validation with repsect to XFoil on an out-of-distribution airfoil. Each colored line represents an analysis at a different Reynolds number. Results are for incompressible, viscous flow with $\Ncr=9$ and natural transition. Solid lines represent NeuralFoil (``NF'') analyses, which are given for two models with different accuracy-speed tradeoffs. The dotted line represents XFoil results at the respective Reynolds number. Agreement with XFoil is almost exact for the ``xxxlarge'' model, across a broad range of angles of attack and Reynolds numbers.}
    \label{fig:clcd_polar}
\end{figure}

The airfoil analyzed here was developed separately for a real-world aircraft development program (\emph{Dawn One}, a high-altitude long-endurance aircraft \cite{sharpeOptimizationApproachMapping2021,sharpeTaileronsAeroelasticStability2023}), using an inverse design process. This airfoil was not included in the training data, either directly or within the ``parent airfoil'' database described in Section \ref{sec:training-data}. Because of this, this represents a previously-unseen airfoil for NeuralFoil, so no unfair advantage is gained by memorization. The airfoil geometry itself is also shown in Figure \ref{fig:clcd_polar}.

In this study, aerodynamic performance is assessed across a range of Reynolds numbers spanning four orders of magnitude. The amplification factor is set as $\Ncr=9$ and natural transition is used. All cases are run in the incompressible limit ($\Mi=0$).

Figure \ref{fig:clcd_polar} shows that excellent agreement is achieved between NeuralFoil and XFoil. The results are nearly exact for the ``xxxlarge'' model, which is the most accurate and computationally expensive model included with NeuralFoil. The ``medium`` model achieves slightly reduced accuracy while offering considerable speed improvements. Table \ref{tab:neuralfoil_performance}.

The Reynolds numbers shown in Figure \ref{fig:clcd_polar} were chosen to illustrate accuracy in a challenging flow condition that occurs near $\Rey=80 \times 10^3$ for this airfoil. Here, the upper-surface boundary layer becomes extremely delicate, as illustrated by the sudden, discontinuous jump near $C_L=1.0$. At angles of attack below this jump, the boundary layer simply undergoes laminar separation and never reattaches. At angles of attack above this jump, a laminar separation bubble (LSB) is formed, which undergoes turbulent reattachment before eventually separating again farther downstream. This specific effect only occurs in a narrow window of airfoil shapes, $\Rey_c$, and $\Ncr$ near the values shown in Figure \ref{fig:clcd_polar}, which forces the network to rely on learned physics. Nevertheless, this phenomenon is well-captured by the ``xxxlarge'' NeuralFoil model, demonstrating its generalization performance. Another benefit of NeuralFoil shown in Figure \ref{fig:clcd_polar} is the inherent $C^\infty$-continuity of its solutions, which smooths out XFoil's ``jagged'' predictions in regions such as the aforementioned discontinuity.

\subsection{Self-Reported Uncertainty Quantification}
\label{sec:uq_results}

\begin{figure}[H]
    \centering
    \includesvg[width=\textwidth]{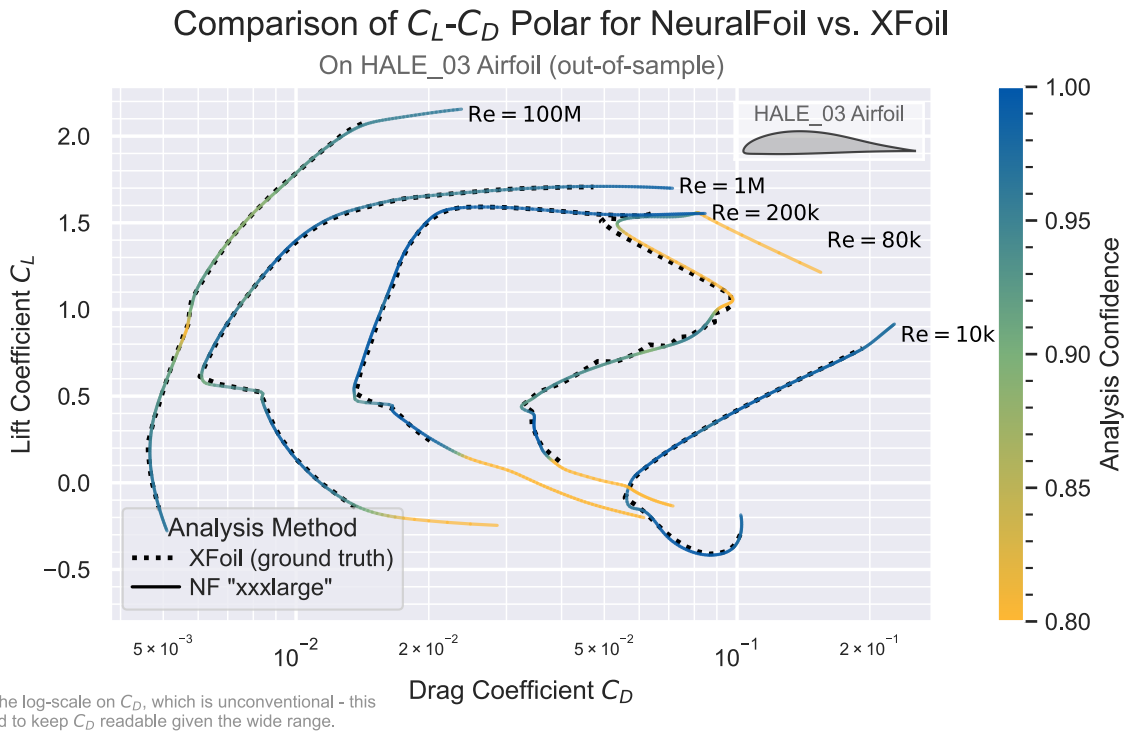}
    \caption{Demonstration of NeuralFoil's self-reported ``analysis confidence'' metric, shown in an identical analysis setup as Figure \ref{fig:clcd_polar}. In this figure, color represents the value of the analysis confidence metric, which varies throughout the input space.}
    \label{fig:clcd_polar_with_uq}
\end{figure}

As discussed in Section \ref{sec:uq}, NeuralFoil self-reports an analysis confidence metric that is intended to aid the user in assessing surrogate trustworthiness. This metric is shown in Figure \ref{fig:clcd_polar_with_uq} for the same airfoil and flow conditions as in Figure \ref{fig:clcd_polar}. Notably, regions with uncertain flow features are immediately flagged. This includes obvious regions of uncertainty, such as post-stall, but also more subtle ones, such as the region near $\Rey_c=80 \times 10^3$ and $C_L=1.0$. Another subtle example is a small region in the $\Rey_c=100 \times 10^6$ case near $C_L=1.0$, where movement of the stagnation point causes sudden changes in the bottom-surface transition location, and thus greater uncertainty.

\subsection{NeuralFoil Accuracy on the Test Dataset}

To report the accuracy of NeuralFoil on a broader range of airfoil shapes and flow conditions, we can measure performance on the test dataset. This dataset was generated using the same methods described in Section \ref{sec:training-data}, but was not used during training. In total, this test dataset consists of 395,665 cases.

The analysis accuracy on this test dataset is shown in Table \ref{tab:neuralfoil_performance}. At a basic level, the figures of merit are accuracy (here, treating XFoil as a ``ground truth'') and computational speed. This table details both considerations. The first set of columns shows the error with respect to XFoil on the test dataset. The second set of columns gives the runtime speed of the models, both for a single analysis and for a large batch analysis. Here, the benefit of NeuralFoil's vectorization is shown: for large batch analyses, runtimes many orders of magnitude faster than XFoil are possible, with minimal loss in accuracy.

\begin{table}[H]
    \begin{centering}
        \caption{Performance comparison of NeuralFoil (``NF'') physics-informed machine learning models versus XFoil in terms of accuracy (treating XFoil as a ground truth) and speed, as measured on the test dataset of Section \ref{sec:training-data}. All NeuralFoil models show significant speed advantages over XFoil, particularly for batch analyses where vectorization provides 100-1,000x speedups. Runtime speeds are measured on an AMD Ryzen 7 5800H laptop CPU.}
        \label{tab:neuralfoil_performance}

        \begin{adjustbox}{width=\textwidth}
            \begin{tblr}{
                    width=\textwidth,
                    colspec={m{2.1cm} | m{2cm} m{2cm} m{2cm} m{2cm} | m{2cm} m{2cm}},
                    row{1}={font=\bfseries},
                }
                \toprule
                Aerodynamics Model & \SetCell[c=4]{c, halign=c, wd=8cm, m} Mean Absolute Error ($L_1$-norm) of Given Metric on the Test Dataset, with respect to XFoil &                                               &                     &                                 & \SetCell[c=2]{c, halign=c, wd=4cm, m} Computational Cost to Run &                              \\
                                   & Lift Coeff. $C_L$                                                                                                                 & Fractional Drag Coeff. $\ln(C_D)\ ^{\dagger}$ & Moment Coeff. $C_M$ & Transition Locations $x_{tr}/c$ & Runtime (1~run)                                                 & Total Runtime (100,000~runs) \\
                \midrule
                NF ``xxsmall''     & 0.040                                                                                                                             & 0.078                                         & 0.007               & 0.044                           & 1.2 ms                                                          & 0.87 sec                     \\
                NF ``xsmall''      & 0.030                                                                                                                             & 0.057                                         & 0.005               & 0.033                           & 1.2 ms                                                          & 1.03 sec                     \\
                NF ``small''       & 0.027                                                                                                                             & 0.050                                         & 0.005               & 0.027                           & 1.3 ms                                                          & 1.14 sec                     \\
                NF ``medium''      & 0.020                                                                                                                             & 0.039                                         & 0.003               & 0.020                           & 1.3 ms                                                          & 1.36 sec                     \\
                NF ``large''       & 0.016                                                                                                                             & 0.030                                         & 0.003               & 0.014                           & 1.3 ms                                                          & 2.34 sec                     \\
                NF ``xlarge''      & 0.013                                                                                                                             & 0.024                                         & 0.002               & 0.010                           & 1.4 ms                                                          & 2.80 sec                     \\
                NF ``xxlarge''     & 0.012                                                                                                                             & 0.022                                         & 0.002               & 0.009                           & 1.6 ms                                                          & 5.13 sec                     \\
                NF ``xxxlarge''    & 0.012                                                                                                                             & 0.020                                         & 0.002               & 0.007                           & 6.1 ms                                                          & 12.0 sec                     \\
                XFoil              & 0                                                                                                                                 & 0                                             & 0                   & 0                               & 73 ms                                                           & 42 min                       \\
                \bottomrule
            \end{tblr}
        \end{adjustbox}
    \end{centering}
    $^{\dagger}$ The deviation of $\ln(C_D)$ can be thought of as ``the typical relative error in $C_D$''. (E.g., $0.020 \rightarrow 2.0\%$ error.) \\
\end{table}

\subsection{Accuracy-Speed Tradeoff vs. XFoil}

When developing machine learning surrogate models based on conventional physics solvers, far too often in the literature a speedup is claimed without controlling for accuracy. After all, it is typically trivial to make the conventional physics solver faster by sacrificing accuracy (e.g., by changing the level of discretization). Therefore, any fair and meaningful speed comparison between a machine learning surrogate and a conventional solver must control for accuracy.

\begin{figure}[H]
    \centering
    \includesvg{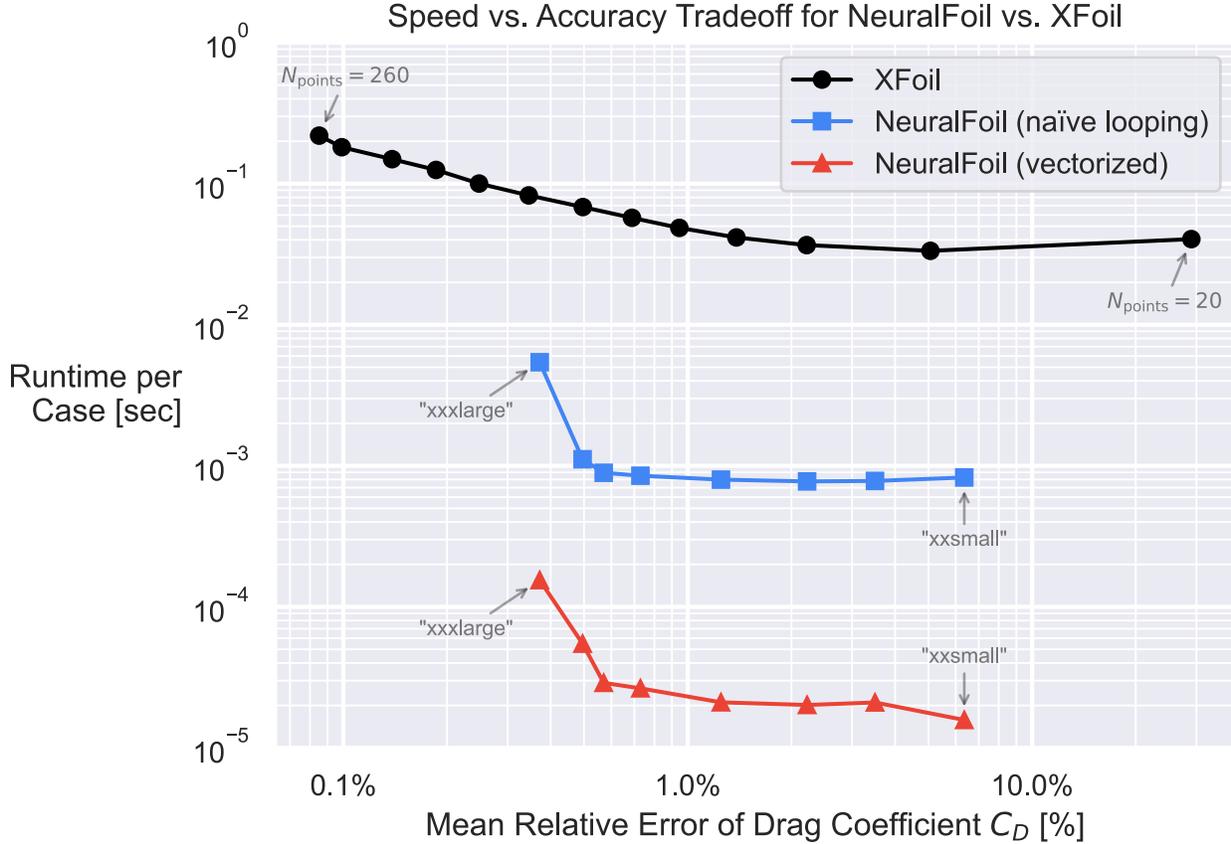}
    \caption{Comparison of runtime speed for NeuralFoil and XFoil, while controlling for accuracy. Evaluated on a varied database of NACA airfoils, with ground truth from XFoil at its highest allowable resolution. XFoil runs are varying resolution; NeuralFoil runs are varying model sizes. Speeds are measured on an AMD Ryzen 7 5800H laptop CPU.}
    \label{fig:accuracy_speed}
\end{figure}

Figure \ref{fig:accuracy_speed} gives this speed-accuracy tradeoff for NeuralFoil, compared to the most common conventional alternative, XFoil. For NeuralFoil, this tradeoff is controlled by the model size (as described in Table \ref{tab:model-sizes}). For XFoil, this tradeoff is controlled by the number of points used to discretize the airfoil, which was varied from 20 to 260 in Figure \ref{fig:accuracy_speed}.

In this study, each model is evaluated on a range of airfoils and flow conditions. To minimize issues with non-convergence in XFoil, a new range of aerodynamic cases is used. The airfoil shapes are NACA 4-series airfoils with the location of maximum camber fixed at $x/c=0.4$. Thickness is varied in the range $[0.08, 0.16]$ and maximum camber is varied in $[0.00, 0.06]$; both are independently varied in increments of $0.01$. This creates a total of $9\times7=63$ airfoil shapes. Each airfoil is evaluated at three Reynolds numbers: $500\times10^3$, $2\times10^6$, and $8\times10^6$. This gives a total of 189 aerodynamic cases. In all cases, $\alpha=5\degree$, $\Mi=0$, and $\Ncr=9$. This selection of aerodynamic cases is relatively ``easy'' and is deliberately chosen to give a runtime speed advantage to XFoil. This is because XFoil, as an iterative algorithm, tends to have an increasing computational cost as flow cases become more difficult. By contrast, NeuralFoil has a fixed computational cost per analysis. Overall, 95\% of cases converged with XFoil, averaged over all discretization levels.

The ``ground truth'' reference solutions for this study are taken from a high-resolution XFoil simulation with 289 points, which is the highest resolution that XFoil 6.98 allows before it begins silently truncating wake points\footnote{This is due to a fixed-size array that XFoil allocates named \texttt{IDX}, which is truncated to avoid overflowing.}.

In Figure \ref{fig:accuracy_speed}, the $x$-axis shows the accuracy of each model, measured as the mean relative error of the drag coefficient\footnote{Or, equivalently, the mean absolute error of $\ln(C_D)$} across all 189 cases. The $y$-axis shows the runtime speed, which is the median time to run each of the 189 cases. By using the median runtime, XFoil again gains a runtime speed advantage, as slow convergence creates a strong positive skew in the distribution of XFoil's runtimes.

As shown in Figure \ref{fig:accuracy_speed}, all of NeuralFoil's model sizes give a speedup over XFoil, even when controlling for accuracy. With naïve looping over all 189 cases, NeuralFoil achieves the same accuracy level as XFoil at speeds more than 8x faster. However, NeuralFoil really shines when taking advantage of its vectorization and analyzing all 189 cases simultaneously. Here, speedups can be nearly 1,000x at the same accuracy level.

Several other notable observations can be made from Figure \ref{fig:accuracy_speed}. First, NeuralFoil's accuracy on ``easy'' cases appears far better than Table \ref{tab:neuralfoil_performance} suggests. This is because the test dataset in Table \ref{tab:neuralfoil_performance} includes many post-stall and transition-sensitive cases, as described in Section \ref{sec:training-data}. For example, the mean relative error of drag for the ``xxxlarge'' model is $2.0\%$ on the test dataset, while the same metric on this easier dataset is $0.37\%$. This reinforces that, when comparing accuracy across studies, the difficulty of the evaluation cases must be considered.

Secondly, NeuralFoil achieves a ``knee'' in the speed-accuracy tradeoff curve roughly near the ``xlarge'' model size. This is therefore taken as the default model size for NeuralFoil, though different use cases may encourage other choices.

\subsection{Sample Validation on Transonic Case}
\label{sec:validation_transonic}

\begin{figure}[H]
    \centering
    \includesvg[width=\textwidth]{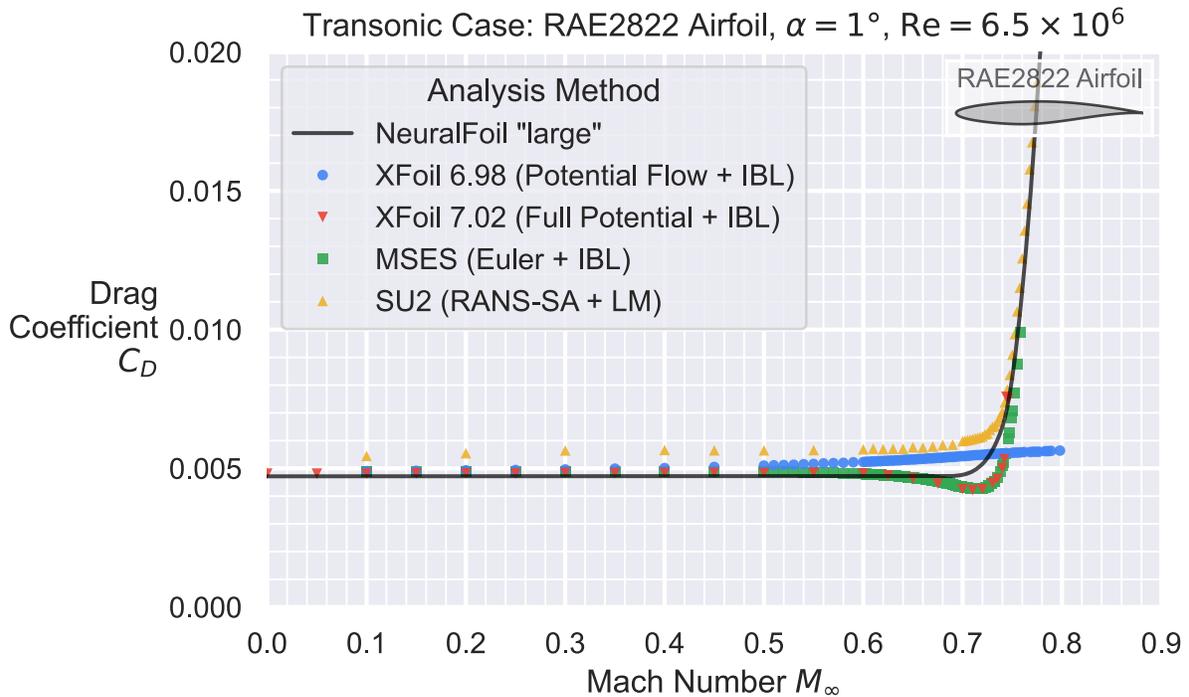}
    \caption{Validation of NeuralFoil at predicting the emergence of transonic flow, using the RAE2822 airfoil at $\Rey_c=6.5 \times 10^6$ and $\alpha=1\degree$. NeuralFoil's results are compared to those from XFoil 6.98, XFoil 7.02, MSES, and SU2. NeuralFoil's results are shown for the ``large'' model described in Table \ref{tab:model-sizes}.}
    \label{fig:transonic_validation}
\end{figure}

To demonstrate the effect of compressibility on NeuralFoil's aerodynamics estimates, we can compare NeuralFoil to various other approaches in a transonic case study. This case study analyzes a RAE2822 supercritical airfoil at flow conditions of $\Rey_c=6.5 \times 10^6$ and $\alpha=1\degree$, with natural transition and $\Ncr=9$. The freestream Mach number is varied from subsonic to transonic conditions. The results of this study are shown in Figure \ref{fig:transonic_validation}. Here, NeuralFoil's results are compared to those from four existing approaches:

\begin{itemize}
    \item XFoil 6.98 \cite{drelaXFOILAnalysisDesign1989}: This uses a boundary-element potential flow solver with a Karman-Tsien compressibility correction for the outer flow, and an integral boundary layer (IBL) model for the boundary layer. Transition is handled with an $e^N$ method.
    \item XFoil 7.02: This uses a grid-based full-potential solver for the outer flow; compared to a linearized potential flow solver, this has stronger theoretical grounding for transonic analysis and the ability to directly capture shock waves. The boundary layer and transition models are the same as those in XFoil 6.98.
    \item MSES \cite{drelaUsersGuideMSES2007}: This is a hybrid Euler / Full-Potential solver that has yet-stronger theoretical grounding as shocks become stronger. The boundary layer and transition models are similar to those in XFoil 6.98, though viscous-inviscid coupling is achieved via a displacement-body approach rather than a wall-transpiration approach. This model is believed to be the most accurate of those listed here for transonic flows with weak shocks.
    \item SU2 \cite{economonSU2OpenSourceSuite2016}: This is a RANS solver with a Spalart-Allmaras turbulence model \cite{spalartOneequationTurbulenceModel1992}. Transition is handled using a Langtry-Menter (LM) correlation-based model \cite{menterCorrelationBasedTransitionModel2006}, using the Malan correlation \cite{menterOneEquationLocalCorrelationBased2015}. Freestream turbulence intensity is set to $0.07\%$, which approximately corresponds to $\Ncr=9$ following equations from Drela \cite{drelaFlightVehicleAerodynamics2013}. However, this is not a direct equivalence, and the LM model predicts slightly earlier transition than $e^{N}$-based methods in this case, causing higher viscous drag.
\end{itemize}

NeuralFoil's results in Figure \ref{fig:transonic_validation} are qualitatively similar to those obtained with other methods, particularly with respect to the location and steepness of the transonic drag rise. Some differences remain, however. Notably, models that use both an integral-boundary-layer approach and a higher-fidelity transonic treatment (i.e., XFoil 7.02, MSES) capture a slight decrease in drag in a window between the critical and drag-divergent Mach numbers. This effect is believed to be real, and it occurs because the locally-supersonic flow in this regime causes a favorable pressure gradient that delays the top-surface boundary layer transition. Such an effect is not captured by NeuralFoil, SU2, or XFoil 6.98, as capturing this effect requires the combination of both shock-resolution and advanced transition modeling.

The results of these analyses can be quantitatively compared in Table \ref{tab:transonic_validation}, which shows the critical and drag-divergent Mach numbers predicted by each method. Corroborating the values in this table, Drela cites the critical Mach number for this case as $\Mcr=0.71$ \cite{drelaFlightVehicleAerodynamics2013}. NeuralFoil appears to underestimate the critical Mach number slightly in this case, although the drag-divergent Mach number agrees very closely with other methods.

\begin{table}[H]

    \begin{centering}
        \caption{Predictions of critical and drag-divergent Mach numbers for the RAE2822 airfoil at $\Rey_c=6.5 \times 10^6$ and $\alpha=1\degree$, using various methods.}
        \label{tab:transonic_validation}
        \begin{tabular}{lll}
            \toprule
                                              & Critical Mach Number $\Mcr$ & Drag-Divergent Mach Number $\Mdd\ ^{\dagger}$ \\ \midrule
            NeuralFoil ``large''              & 0.671                       & 0.739                                         \\
            NeuralFoil ``xxxlarge''           & 0.673                       & 0.741                                         \\
            XFoil 6.98 (Potential Flow + IBL) & 0.688                       & N/A                                           \\
            XFoil 7.02 (Full Potential + IBL) & 0.705                       & 0.739                                         \\
            MSES (Euler + IBL)                & 0.714                       & 0.738                                         \\
            SU2 (RANS-SA + LM)                & -                           & 0.738                                         \\ \bottomrule
        \end{tabular}
    \end{centering}
    \\
    $^{\dagger}$ Defined as $\partial(C_D)/\partial \Mi \geq 0.1$, following Mason \cite{masonTransonicAerodynamicsAirfoils2006}.\\

\end{table}

\subsection{Comparison to other Airfoil Machine Learning Models}

Here, we compare and contrast NeuralFoil with several prior attempts to apply machine learning techniques to airfoil aerodynamics analysis. Peng et al. \cite{pengLearningAerodynamicsNeural2022} apply a convolutional neural network to this task, where the input is an array of airfoil coordinates and the output is a scalar lift coefficient. Bouhlel et al. \cite{bouhlelScalableGradientEnhanced2020} use feedforward neural networks to predict airfoil lift and drag coefficients; this is augmented by Sobolev regularization that adds gradient information into the loss function. Du et al. \cite{duRapidAirfoilDesign2021} use multilayer perceptrons, recurrent neural networks, and mixture-of-experts approaches to predict both scalar (lift and drag) and vector (pressure distribution) quantities.

Compared to prior work, NeuralFoil is trained on a significantly broader input space, particularly with respect to angle of attack and transition assumptions. This comparison is illustrated in Table \ref{tab:ml_comparison}. The accuracy values listed in Table \ref{tab:ml_comparison} are not directly comparable, as they are computed on different datasets and with different metrics. Nevertheless, NeuralFoil achieves similar accuracy to prior work, despite including many extreme cases in its evaluation sample (e.g., post-stall flows, transitional Reynolds numbers, and high-curvature airfoils associated with 18 shape variables). This is possible partially because of the much larger volume of training data, but also due to the structural embedding of physics knowledge into the model described throughout Section \ref{sec:methodology}.

Accurate and rapid aerodynamics analysis across this broader input space has the potential to enable airfoil design optimization without overly restricting the design space. Furthermore, certain types of robust optimization become possible. For example, performing a multipoint optimization study with respect to $\Ncr$ becomes possible with NeuralFoil, allowing the designer to mitigate the tendency of optimizers to ``over-optimize'' to one specific transition location\footnote{For more details on this, see Drela \cite{drelaProsConsAirfoil1998}. In general, airfoil shape optimization algorithms tend to geometrically ``fill in'' locations where laminar separation bubbles would otherwise go, which dramatically worsens off-design performance.}.

\begin{table}[H]
    \begin{centering}
        \caption{Comparison of NeuralFoil to prior literature in airfoil aerodynamics machine learning. NeuralFoil achieves similar accuracy, despite including a much broader range of flow conditions in its training and evaluation distributions. Models from Du et al. \cite{duRapidAirfoilDesign2021}, Bouhlel et al. \cite{bouhlelScalableGradientEnhanced2020}, and Peng et al. \cite{pengLearningAerodynamicsNeural2022} use separate subsonic and transonic models.}
        \label{tab:ml_comparison}

        \begin{adjustbox}{width=\textwidth}
            \begin{tblr}{
                width=\textwidth,
                colspec={@{} m{2.5cm} m{2.5cm} m{3.1cm} m{3.1cm} m{3.1cm} m{3cm}@{}},
                row{1}={font=\bfseries},
                column{2}={font=\bfseries},
                cell{2-4}{3}={bg=g},
                cell{5}{3}={bg=m},
                cell{6}{3}={bg=g},
                cell{2}{4}={bg=g},
                cell{3}{4}={bg=b},
                cell{4}{4}={bg=g},
                cell{5}{4}={bg=g},
                cell{6}{4}={bg=b},
                cell{2}{5}={bg=m},
                cell{3}{5}={bg=m},
                cell{4}{5}={bg=b},
                cell{5}{5}={bg=g},
                cell{6}{5}={bg=b},
                cell{2}{6}={bg=b},
                cell{3}{6}={bg=m},
                cell{4-6}{6}={bg=b},
                }
                \toprule
                                                                                                                    &                                          & NeuralFoil                                                                             & Du et al. \cite{duRapidAirfoilDesign2021}           & Bouhlel et al. \cite{bouhlelScalableGradientEnhanced2020} & Peng et al. \cite{pengLearningAerodynamicsNeural2022}   \\ \midrule
                \SetCell[r=6]{m,2.5cm}{Diversity of ``airfoil design space'' used for both training and evaluation} & Airfoil shape diversity                  & 18 shape variables                                                                     & 16 movable control points                & {14 shape vars. (subsonic)                                                            \\8 shape vars. (transonic)} & UIUC database ($\sim$1500 fixed shapes) \\
                                                                                                                    & Angle of attack range                    & Uniform + normal distrib.; central 95\% in range $[-17\degree,\ +18\degree]$           & $0\degree$ to $3\degree$                 & {$-0.5\degree$ to $6\degree$                                                          \\(subsonic) \\$-1.5\degree$ to $4.5\degree$\\(transonic)} & $-2\degree$ to $10\degree$ \\
                                                                                                                    & Reynolds number range                    & Log-normal distrib.; central 95\% in range $[1.9\times 10^3,\ 262\times10^6]$          & {$10^4$ to $10^{10}$ (training)}         & None specified                              & $13\times 10^6$ (fixed)                 \\
                                                                                                                    & Mach number range                        & 0 for learned core, although compressibility correction allows prediction up to $\Mcr$ & {0.3 to 0.6 (subsonic)                                                                                                           \\ 0.6 to 0.7 (transonic)} & {0.3 to 0.6 (subsonic) \\0.7 to 0.75 (transonic)} & 0.3 (fixed) \\
                                                                                                                    & Transition, turbulence, and forced trips & $\Ncr$ varied from 0 to 18; Trips varied from LE to TE                                 & None specified ($\tilde{\nu}=0$ assumed) & None specified ($\tilde{\nu}=0$ assumed)    & None specified ($\Ncr=9$ assumed)     & \\
                                                                                                                    & Training data                            & {XFoil                                                                                                                                                                                                                    \\(7.5M samples)} & {RANS-SA via ADflow\\(86k samples)} & {RANS-SA via ADflow\\(42k samples)} & {XFoil\\(16k samples)} \\ \midrule
                \SetCell[r=2]{m,2.5cm}{Performance, as evaluated on the corresponding distribution above}           & $C_L$ mean absolute error                & 0.012                                                                                  & 0.010                                    & Unspecified                                 & 1.0\%                                   \\
                                                                                                                    & $C_D$ mean relative error                & 2.0\%                                                                                  & 1.7\%                                    & 0.3\%                                       & N/A (not attempted)                     \\
                \bottomrule
            \end{tblr}
        \end{adjustbox}
    \end{centering}
\end{table}

\section{Airfoil Design Optimization with NeuralFoil}
\label{sec:optimization}

Although the accuracy of NeuralFoil with respect to XFoil has been demonstrated in Section \ref{sec:results}, this alone is not sufficient to demonstrate that NeuralFoil is useful for design optimization. A major reason for this is that engineering design optimization is an adversarial process, where the optimizer aims to exploit not only the underlying physics, but also any errors in the model. Exploitation of the physics is desirable and ultimately the goal of engineering design optimization, but exploitation of model errors leads to designs that are ostensibly promising but lose their luster when analyzed with other computational tools or actually built and flown.

\begin{figure}[h]
    \centering
    \includesvg[width=\textwidth]{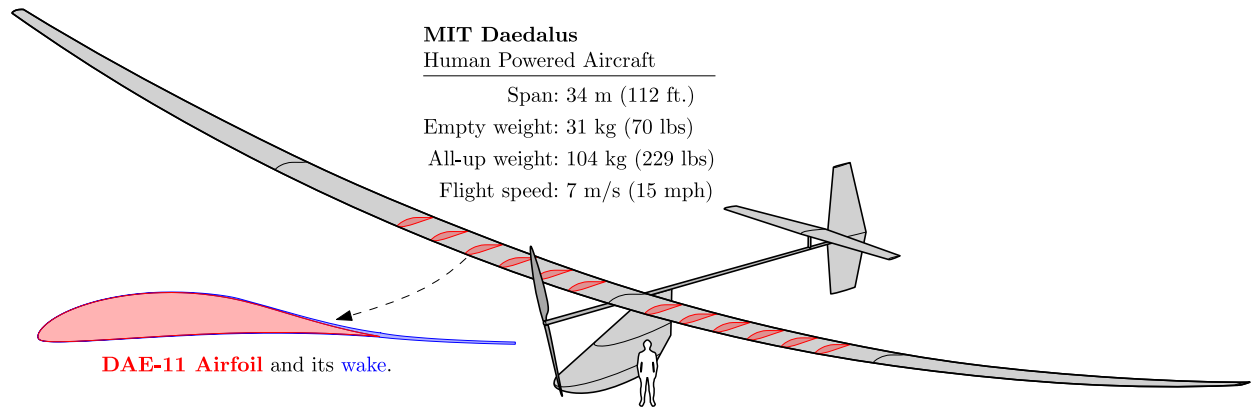}
    \caption{Drawing of the \emph{MIT Daedalus} human-powered aircraft, along with its centerline airfoil, the DAE-11. This airfoil is the subject of the airfoil design optimization problem posed in this section.}
    \label{fig:daedalus_iso}
\end{figure}

To assess whether a model such as NeuralFoil is prone to model error exploitation, we set up an aerodynamic shape optimization problem with a known answer from the literature, and evaluate how close the optimized result matches this. In aerodynamic shape optimization (and aircraft design more broadly), there are relatively few problems that are rigorously specified, and even fewer where the correct optimized result is provided in exacting detail. However, one such case study that provides a useful design problem is Drela's 1998 \textit{Pros \& Cons of Airfoil Optimization} \cite{drelaProsConsAirfoil1998}, which discusses airfoil design for the \emph{MIT Daedalus} human-powered aircraft \cite{langfordFeasibilityHumanPoweredFlight1986, langfordDaedalusProjectSummary1989,drelaHumanPoweredFlight1985}. Basic figures for this aircraft, along with its centerline airfoil, the DAE-11, are shown in Figure \ref{fig:daedalus_iso}.

The design of the DAE-11 airfoil is effectively defined by the following airfoil design optimization problem; this formulation is taken directly from Drela \cite{drelaProsConsAirfoil1998} with minor modifications\footnote{Specifically, a constant-$\Rey_c \sqrt{C_L}$ (fixed-lift) polar is used instead of a constant-$\Rey_c$ polar; the $C_M$ constraint is clarified to apply to all $C_L$ operating points; and the $\theta_{\rm TE}$ constraint is matched to the DAE-11 airfoil}:

\begin{itemize}
    \item \textbf{Objective}: Minimize $C_{D, \mathrm{mean}}$ at $\mathrm{Re}_c = 500\mathrm{k} \cdot \left(\frac{C_L}{1.25}\right)^{-0.5}$ and $\mathrm{M}_\infty = 0.03$
          \begin{itemize}
              \item Where $C_{D, \mathrm{mean}}$ is the weighted average of $C_D$ values at $C_L = [0.8, 1.0, 1.2, 1.4, 1.5, 1.6]$, with relative weights of $[5, 6, 7, 8, 9, 10]$ at each respective $C_L$
          \end{itemize}
    \item \textbf{Variables}: airfoil shape (18 variables; described in Section \ref{sec:airfoil-parameterization}) and angle of attack $\alpha$
    \item \textbf{Constraints}:
          \begin{itemize}
              \item $C_M \geq -0.133$ (at all $C_L$ operating points)
              \item Trailing-edge angle $\theta_{\rm TE} \geq 6.03^\circ$ (allows manufacturability)
              \item Leading-edge angle $\theta_{\rm LE} = 180^\circ$ (prevents the formation of a sharp leading edge)
              \item At $x/c = 0.33$, $t/c \geq 0.128$ (to accommodate the main spar)
              \item At $x/c = 0.90$, $t/c \geq 0.014$ (to accommodate the rear spar)
          \end{itemize}
\end{itemize}

When posing airfoil design optimization problems (using NeuralFoil or with any other tool), it is often helpful to add basic regularization constraints; these tend to make airfoil optimization better-behaved by ruling out non-physical and unrealistic airfoils. For the simple airfoil design problem described above, these regularization strategies are not necessary for convergence. However, they are discussed here as ``best practices'' for airfoil optimization on more complicated problems:

The first such constraint is that the airfoil thickness must be positive everywhere. This prevents the creation of self-intersecting airfoils, which can be analyzed in both XFoil and NeuralFoil but are clearly not physically-valid shapes.

The second such constraint aims to guide the optimizer away from airfoils that have excessively high local surface curvature, as these have boundary layer behavior that is inherently difficult to characterize. A useful global metric for this curvature, which we call the wiggliness $w$, is mathematically similar to a metric by Wahba \cite{wahbaSplineModelsObservational2007} that has long proven successful in univariate spline regularization:

$$w(\mathrm{airfoil}) = \int_0^1 \left( \frac{d^2}{dx^2} y_\mathrm{lower}(x) \right)^2 + \left( \frac{d^2}{dx^2} y_\mathrm{upper}(x) \right)^2 dx$$

\noindent We can loosely approximate the spirit of this wiggliness measure from the discrete Kulfan parameterization as follows:

$$w(\mathrm{airfoil}) \approx \sum_{i=2}^{N-1} \left(c_{\mathrm{lower},i+1} - 2 \cdot c_{\mathrm{lower},i} + c_{\mathrm{lower},i-1} \right)^2 + \left(c_{\mathrm{upper},i+1} - 2 \cdot c_{\mathrm{upper},i} + c_{\mathrm{upper},i-1} \right)^2$$

\noindent where $c_{\mathrm{lower},i}$ and $c_{\mathrm{upper},i}$ are the $i$th Kulfan (CST) coefficients of the lower and upper surfaces, respectively. A reasonable heuristic for this regularization constraint is to restrict this metric to no more than four times that of the original initial guess airfoil (which is typically some generic airfoil, such as a NACA0012).

Using the problem formulation described above, we can solve this airfoil design optimization problem by coupling NeuralFoil with the AeroSandbox aircraft design optimization framework \cite{sharpeAeroSandboxDifferentiableFramework2021}. This adds automatic differentiation capabilities to NeuralFoil, allowing efficient optimization using gradient-based methods. This optimization problem is solved in approximately 7 seconds on a standard laptop; the speed of this solution provides rapid feedback to the designer on how to improve the optimization problem formulation to capture design intent.

\begin{figure}[H]
    \centering
    \includesvg[width=\textwidth]{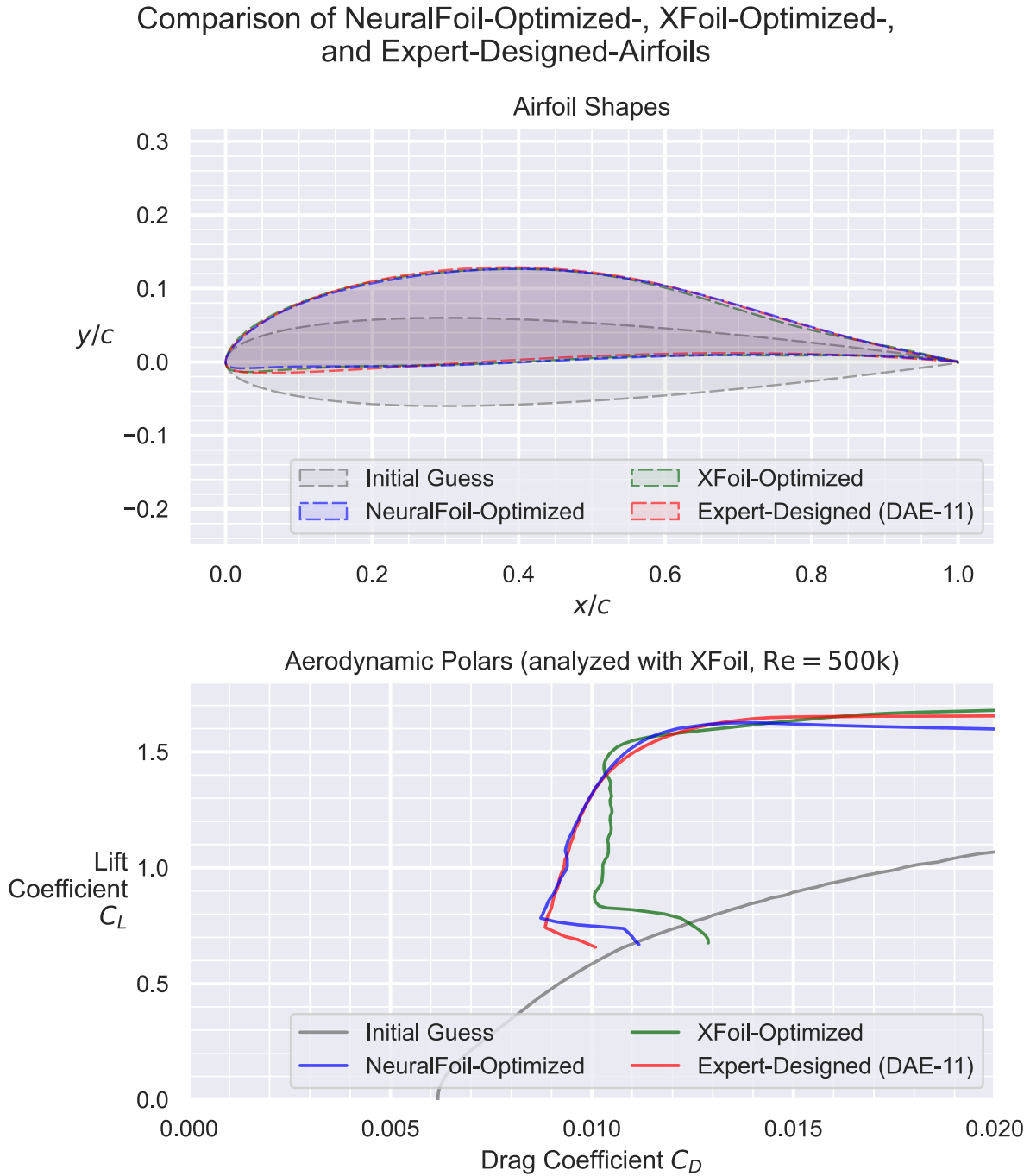}
    \caption{NeuralFoil-optimized airfoils have similar performance and shape to expert-designed airfoils. Adversarial optimization is not observed, as the NeuralFoil-optimized and XFoil-optimized airfoils yield similar performance when analyzed using XFoil. The gray ``Initial Guess'' airfoil is used to initialize NeuralFoil optimization.}
    \label{fig:daedalus_optimized}
\end{figure}

Figure \ref{fig:daedalus_optimized} compares three airfoil designs produced with different methodologies, each aimed at solving the \emph{MIT Daedalus} airfoil design problem described previously:

\begin{itemize}
    \item \textbf{NeuralFoil-optimized}, which is the result of optimizing while using NeuralFoil as the aerodynamics analysis tool
    \item \textbf{XFoil-optimized}, which is the result of optimizing while using XFoil as the aerodynamics analysis tool
    \item \textbf{Expert-designed}, which is the original DAE-11 airfoil designed by Drela \cite{drelaLowReynoldsnumberAirfoilDesign1988} and used on the as-flown \emph{MIT Daedalus} aircraft
\end{itemize}

For comparison, Figure \ref{fig:daedalus_optimized} also includes the initial guess airfoil that was provided to the optimization algorithms (a simple NACA0012). This showcases that the optimization process is robust to initial guesses that are relatively far from the optimal solution.

The airfoil produced using NeuralFoil optimization is quite similar in shape to the those produced using XFoil optimization and expert-designed airfoils, when given the same design objectives and constraints. Likewise, aerodynamic performance is quite similar across all three airfoils. Notably, the aerodynamic polars depicted in Figure \ref{fig:daedalus_optimized} were produced by post-optimality analysis using XFoil. Given the minimal discrepancy between the NeuralFoil-optimized and XFoil-optimized airfoils, this suggests that NeuralFoil is reasonably resistant to model error exploitation on problems of practical interest.

Interestingly, the geometric differences between the expert-designed and optimized airfoils (both from XFoil and NeuralFoil) in Figure \ref{fig:daedalus_optimized} are attributable to design goals that were not factored into the quantitative problem formulation. For example, the optimized airfoils both exhibit a small amount of lower-surface concavity in the vicinity of $x/c \approx 0.15$, while the expert-designed DAE-11 has a flatter lower surface. Drela discusses reasons for this discrepancy in \cite{drelaProsConsAirfoil1998}, noting that the concavity that the optimizer prefers would cause the wing covering (in the case of \emph{MIT Daedalus}, a thin shrunk covering of Mylar) to lift off of the surface of the rib, creating a bubble. This undesirable effect is not captured in the quantitative problem formulation, providing a cautionary tale that an optimized airfoil is only as good as the problem formulation that led to it.

Nevertheless, the strength of NeuralFoil-powered optimization is that it allows one to generate optimized airfoils that are remarkably similar to expert-designed airfoils in mere seconds. This optimization capability is more-than-adequate for conceptual aircraft design, and it provides an excellent starting point for expert-guided airfoil design refinement.

\section{Computational Reproducibility Statement}
\label{sec:reproducibility}

NeuralFoil is implemented as an open-source Python package with minimal dependencies (only NumPy \cite{harrisArrayProgrammingNumPy2020} for the actual aerodynamic modeling), allowing easy installation across a variety of platforms.

All source code used in this paper is publicly available at \url{https://github.com/peterdsharpe/neuralfoil}. Model architecture, weights, training scripts, and usage examples are included in this repository.
For end-users, NeuralFoil is best accessed through the open-source AeroSandbox aircraft design optimization framework \cite{sharpeAeroSandboxDifferentiableFramework2021}; this provides some of the advanced features (e.g., $360\degree$ angle of attack, compressibility corrections, and control surface deflections) described in this work that are not yet available in the standalone NeuralFoil package.

\section{Conclusion}
\label{sec:conclusion}

In this work, we introduce NeuralFoil, a physics-informed machine learning tool for airfoil aerodynamics analysis. Relative to existing popular tools such as XFoil, NeuralFoil offers improved computational efficiency, even after controlling for equivalent accuracy. Its ability to handle a wide range of practical airfoil shapes and flow conditions makes it a versatile tool in aerodynamic analysis.

The accuracy of NeuralFoil with respect to XFoil is demonstrated across several test cases, with a particular focus on airfoils that NeuralFoil was not trained on (i.e., out-of-distribution). Much of the accuracy achieved here is due to the embedding of domain-specific physics knowledge into the model architecture, which increases the parameter-efficiency and generalizability of the resulting model.

Also presented in this work is an application of NeuralFoil to airfoil design optimization for the \emph{MIT Daedalus} human-powered aircraft, where it quickly produces designs close in performance and shape to expert-crafted airfoils. Incorporation of geometric constraints into this study shows that NeuralFoil-based optimization is able to capture the non-aerodynamic considerations that are typical of practical design problems. This case study also demonstrates the real-world value of many optimization-friendly features that are embedded into the model architecture (e.g., continuity, differentiability, and static code execution paths). Finally, these results also suggest that NeuralFoil is not prone to adversarial exploitation of model errors.

Another contribution of this work is a new technique that enables machine learning surrogate models to estimate their own trustworthiness, a critical capability for robust engineering design. This ``analysis confidence'' metric can be used directly to guide human-in-the-loop analysis, or it can be directly constrained during a formal optimization process to improve the robustness of the resulting designs.

Future work could include extending NeuralFoil to handle multi-element airfoils, which would require revisiting the geometry parameterization among other considerations. Another useful research direction would be to include compressibility corrections directly within NeuralFoil's learned model, as opposed to using an analytical correction to incompressible results. Currently, the outer flow receives a compressibility correction, but the effects of this modified pressure distribution are not then fed back into the boundary layer model. Therefore, some transonic boundary layer physics (such as an extended laminar run within a supersonic zone that causes a favorable pressure gradient) are fundamentally lost. This does, however, increase the dimensionality of the input space, and thus may affect training data requirements. Related to that, another possible next study could investigate how model accuracy changes with the amount of training data used. NeuralFoil was trained with roughly two orders of magnitude more aerodynamic cases than similar studies, although it is not yet known whether this is strictly necessary to achieve the model's performance.

As a Python-based, open-source package, NeuralFoil offers a practical and versatile solution that is accessible to both academia and industry. Its accurate and rapid performance offer value throughout the aircraft development process, though this capability is especially interesting in the early stages of aircraft development where quick feedback on design changes is crucial.

\section*{Acknowledgments}
The authors acknowledge the MIT SuperCloud and Lincoln Laboratory Supercomputing Center for providing high-performance computing resources that have contributed to the research results reported within this work.

\bibliography{main, C:/Users/psharpe/library}

\end{document}